\documentclass[journal=jacsat,manuscript=article,layout=standard]{achemso}

\usepackage{tabularx}
\usepackage{booktabs}
\usepackage{setspace}
\usepackage{xcolor}
\usepackage{multirow}
\usepackage{graphicx}
\usepackage[version=4]{mhchem}
\usepackage{pdfpages}

\author{Dedi Sutarma}
\author{Peter Kratzer}
\email{dedi.sutarma@uni-due.de}
\affiliation{Department of Physics, University of Duisburg-Essen, Duisburg}

\title[An \textsf{achemso} demo]
  {Facet-Dependent Electronic Properties and Interfacial Point Defect Interactions in WS$_2$/ZnO Heterostructures}


\begin{document}
\singlespacing

%\begin{tocentry}

%\end{tocentry}

\begin{abstract}
Aiming at the use of two-dimensional materials as components in high-efficiency optoelectronics, WS$_2$/ZnO heterostructures are computationally screened for their facet-dependent electronic properties and interfacial defect thermodynamics by means of first-principles calculations with a hybrid density functional. Comparison of our interface models identifies the non-polar ($10\overline{1}0$) $m$-plane as the optimal substrate facet, maintaining a direct bandgap of 2.42~eV and a robust type-I band alignment necessary for carrier confinement. Our study on isolated point defects reveals that sulfur ($\mathrm{V}_{\mathrm{S}}$) and interfacial oxygen ($\mathrm{V}_{\mathrm{O}}$) vacancies introduce deep-level states that act as non-radiative recombination centers. Conversely, zinc vacancies ($\mathrm{V}_{\mathrm{Zn}}$) act as shallow acceptors near the valence band edge, contributing to unintentional $p$-type behavior due to the Type-I band alignment. Analysis of defect pairs shows that neutral vacancies cluster across the van der Waals gap due to favorable negative binding energies. Near the conduction band minimum ($n$-type conditions), charge transition level analysis indicates that defects stabilize as charged species. Although inter-layer Coulomb repulsion weakens the binding energy of $(\mathrm{V}_{\mathrm{S}} - \mathrm{V}_{\mathrm{Zn}})''''$ pairs, their formation energy drops to 2.61~eV under anion-poor conditions, making the $-4$ cluster the most thermodynamically abundant defect pair at the interface. Furthermore, the native zinc vacancy actively prevents the Fermi level rise typically induced by interstitial hydrogen; in the $\mathrm{H}_{\mathrm{i}}+\mathrm{V}_{\mathrm{Zn}}$ pairs, the donated electron is distributed into shallow acceptor states, preserving the rigidity of the host band edges. These findings establish a microscopic framework for substrate selection and defect engineering to optimize the internal quantum efficiency of 2D/3D hybrid light-emitting diodes.
\end{abstract}

\section{Introduction}
The integration of two-dimensional (2D) transition metal dichalcogenides (TMDs) with three-dimensional (3D) wide-bandgap semiconductors has emerged as an attractive strategy for advancing optoelectronic performance, particularly in the field of light-emitting diodes (LEDs)\cite{andrzejewski_scalable_2019, andrzejewski_flexible_2020}. Among these hybrid systems, the heterostructure formed by monolayer tungsten disulfide (WS$_2$) and zinc oxide (ZnO) offers a unique materials platform defined by favorable excitonic stability and band energetics\cite{janotti_fundamentals_2009,kim_biexciton_2016}. While ZnO is a well-established wide-bandgap semiconductor ($E_g \approx 3.37$ eV) with a large exciton binding energy of 60~meV -- significantly higher than the 25~meV observed in traditional III-V materials like GaN -- monolayer WS$_2$ represents a paradigm shift in excitonic physics\cite{janotti_native_2007,janotti_fundamentals_2009}. 
Upon thinning to the monolayer limit, WS$_2$ transitions to a direct bandgap semiconductor with an high exciton binding energy exceeding 500~meV. This value is an order of magnitude larger than that of bulk semiconductors, ensuring that excitonic states in WS$_2$ remain thermally stable well above room temperature, thereby resisting dissociation into free carriers and enabling robust radiative recombination efficiencies essential for high-performance LEDs\cite{kim_biexciton_2016-2}.

Efficient carrier transport and injection mechanics are important for the operation of these 2D/3D heterojunction LEDs. The band offsets at the WS$_2$/ZnO van der Waals interface play a decisive role in transporting electrons and holes into the WS$_2$ active layer\cite{butanovs_fast-response_2018}. Specifically, in a sandwiched configuration (e.g., ZnO/WS$_2$/ZnO), the band alignment promotes the transfer of photogenerated or electrically injected carriers from the high-energy states in the ZnO cladding layers down into the conduction and valence bands of the WS$_2$ monolayer\cite{bora_manipulating_2024}. This mechanism ensures that both electrons and holes accumulate within the 2D channel, maximizing the probability of radiative recombination\cite{butanovs_fast-response_2018}. Because the van der Waals interface lacks dangling bonds, it exhibits a low density of non-radiative recombination centers relative to conventional covalent heterojunctions, preserving high internal quantum efficiency.\cite{pham_2d_2022}.

The crystallographic orientation of the zinc oxide (ZnO) substrate functions as an important determinant for the structural integrity and electronic quality of the interface in WS$_2$/ZnO heterostructures. Although bulk wurtzite ZnO exhibits hexagonal symmetry, the surface energy and chemical reactivity governing 2D layer nucleation differ between the polar $c$-plane and non-polar $m$- and $a$-planes. To neutralize polarity, the (0001) facet can undergo massive reconstruction, involving the formation of zinc vacancies (V$_{Zn}$) or the adsorption of hydroxyl groups\cite{mora-fonz_why_2017-1, heinhold2013influence}. Consequently, while the high surface energy of the $c$-plane\cite{wang_vacancy_2022} might promote strong adhesion, the interface is inherently disordered and non-stoichiometric, often presenting triangular pits or step edges that can act as scattering centers for carriers in the overlying WS$_2$ layer.

In contrast, the non-polar $(10\bar{1}0)$ $m$-plane and $(11\bar{2}0)$ $a$-plane facets consist of equal numbers of cations and anions within the surface plane, eliminating the spontaneous polarization and the need for charge-compensating defects. Although non-polar ZnO facets present a structurally passivated substrate for WS$_2$ growth, native point defects in both the $2\text{D}$ monolayer and the 3D oxide introduce localized electronic states that dictate the heterostructure's transport and optical properties.

While the individual defect chemistries of monolayer tungsten disulfide (WS$_2$) and bulk zinc oxide (ZnO) have been studied considerably, a critical knowledge gap remains regarding their cooperative behavior at the interface. Extensive first-principles studies have characterized native point defects in ZnO, identifying the oxygen vacancy (V$_O$) as a deep donor and the zinc vacancy (V$_{Zn}$) as a dominant deep acceptor responsible for compensation mechanisms\cite{oba_point_2011,oba_defect_2008,knutsen_zinc_2012}. Similarly, the sulfur vacancy (V$_S$) in WS$_2$ has been identified as a source of deep in-gap states that can trap charge carriers and induce non-radiative recombination\cite{schuler_large_2019, haldar_systematic_2015}.

We argue that these point defects do not strictly act in isolation. Due to the reduced coordination at the van der Waals interface, the formation energy threshold for native defects is significantly lowered, leading to a dense accumulation of isolated vacancies with large capture cross-sections for charge carriers. 
In the context of Shockley-Read-Hall (SRH) statistics, these deep-level clusters act as potent non-radiative recombination centers. The multiphonon emission processes associated with carrier capture at these deep levels dissipate energy as heat rather than light, drastically quenching the internal quantum efficiency of the LED\cite{bora_manipulating_2024,kangsabanik_defect-assisted_2026}.

This work utilizes density functional theory (DFT) to evaluate the influence of ZnO crystallographic orientation---specifically comparing the polar (0001) $c$-plane with the non-polar $(10\bar{1}0)$ $m$-plane and $(11\bar{2}0)$ $a$-plane---to determine how surface-specific strain and electrostatic potentials modulate the heterostructure properties. Building upon this, we provide the comprehensive identification of the defect pairs that are thermodynamically favored at the interface. By establishing the thermodynamic driving forces that govern this interfacial defect accumulation, we elucidate the microscopic origins of efficiency droop in 2D/3D optoelectronics and define defect-engineering strategies required to restore ideal Type-I performance.

\section{Results and discussion}

\subsection{Facet-Dependent Structural Stability and Electronic Coupling}

The structural integrity and electronic performance of WS$_2$/ZnO heterojunctions are strongly influenced by the crystallographic orientation of the ZnO substrate. We performed a comparative analysis of monolayer WS$_2$ on the non-polar $(10\bar{1}0)$ $m$-plane, the non-polar $(11\bar{2}0)$ $a$-plane, and the hydroxyl-passivated polar $(0001)$ $c$-plane. 

Structural models for all three interfaces were built using supercells that are a compromise between minimization of strain and computational feasibility. In experimental van der Waals heterostructures, the relatively weak interlayer coupling often allows the WS$_2$ layer to maintain its intrinsic lattice constant through an incommensurate arrangement. The calculated structural and electronic parameters for these models are summarized in Table~\ref{tbl:facets}. Additionally, it should be noted that the $(11\bar{2}0)$ $a$-plane introduces strong anisotropy. The lattice relaxation on the $a$-plane is distinct in the $x$ and $y$ directions, creating an anisotropic strain environment for the 2D material. The higher energy of the $a$-plane compared to the $m$-plane, combined with its anisotropic lattice structure, also suggests different adhesion mechanics for 2D layers. 

The transition from the free-standing WS$_2$ to the heterostructure interface induces a shift of the electronic bandgap, primarily by interfacial strain. In addition, the screening of the long-ranged exchange and correlation effects due to the presence of a substrate can reduce the bandgap. For comparison, we note that free-standing monolayer WS$_2$ possesses a bandgap of 2.51~eV in the computational approach used here.  

For assessing the effect of strain, the band structure of free-standing WS$_2$ layers strained to match various substrates are displayed in Fig.~S1. 

\begin{figure}[ht!]
     \centering
     \includegraphics[width=0.95\textwidth]{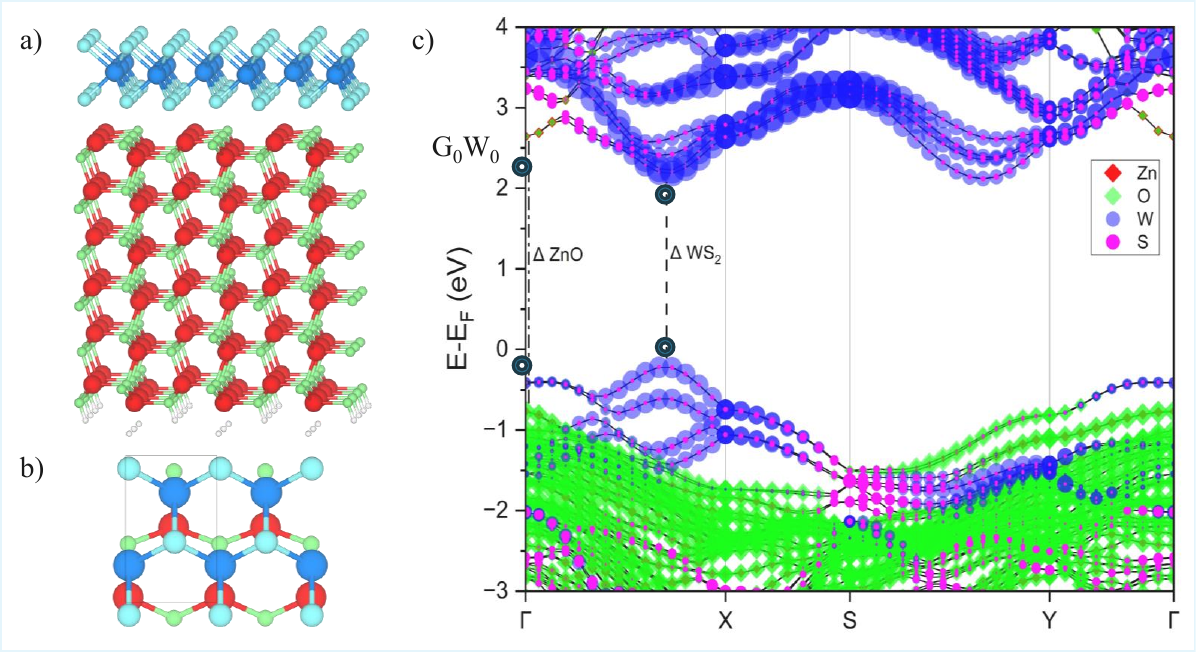}
     \caption{Structural and electronic characterization of the $WS_2/ZnO(10\bar{1}0)$ $m$-plane heterostructure. a) side-view supercell, b) top-view illustrating the near-eclipsed stacking, and c) element-resolved band structure. The interface preserves the direct bandgap of WS$_2$ and maintains a robust Type-I alignment.}
     \label{fig:m_plane}
\end{figure}

At the $(10\bar{1}0)$ $m$-plane interface, the ZnO gap found to be 3.42~eV, while the WS$_2$ gap is reduced to 2.41~eV. Crucially, the $m$-plane preserves the direct bandgap nature of WS$_2$ (see Fig.~\ref{fig:m_plane}a) despite a compressive strain energy of 9~meV/\AA$^2$. As seen in the top view of Fig.~\ref{fig:m_plane}b, the chemical bonds at the $m$-plane interface nearly adopt a an eclipsed configuration, characterized by vertical alignment of $W - S$ bonds almost directly above $Zn - O$ bonds. This model, characterized by moderate interfacial distance ($2.42$~\AA), effectively balances interlayer repulsion and attractive forces, maintaining the direct bandgap of WS$_2$. This configuration achieves a robust type-I alignment where the 2D gap is nested within the wider ZnO gap. 

The $K$ point, which forms the valence band maximum and conduction band minimum in isolated monolayer WS$_2$, folds back onto the $\Gamma\text{--}X$ path within the folded supercell Brillouin zone. The valence band maximum (VBM) of WS$_2$ comes to lie 0.55~eV above the VBM of ZnO, while the conduction band minimum (CBM) of WS$_2$ is 0.46~eV below the CBM of ZnO. To ensure the accuracy of these alignments, our hybrid functional calculations ($\alpha = 0.375$) for WS$_2$/ZnO$(1 \bar 1 00)$ were benchmarked against $G_0W_0$ results. It was found to still underestimate the $Zn$ 3d state positioning and the overall band gaps (2.9~eV for ZnO and 2.3~eV for WS$_2$), which was observed in previous studies\cite{usuda2002all, shishkin2007self, fuchs2007quasiparticle}

To further evaluate the impact of the dielectric environment, we investigated a double-sided 'sandwich' model (ZnO/WS$_2$/ZnO) on the $m$-plane (Fig. S4). This architecture mirrors encapsulated device designs where 2D layers are protected by top and bottom dielectric barriers\cite{bora_manipulating_2024}. In this configuration, the lattice mismatch remains constant at 2.67\%, while the WS$_2$ bandgap persists at 2.39~eV. 

As demonstrated in literature for MoS$_2$ systems\cite{ryou_monolayer_2016}, surrounding a 2D semiconductor with a high-$\kappa$ medium provides additional screening that stabilizes the quasiparticle gap against external fluctuations. Our results confirm that the type-I alignment is preserved in this environment with minimum change to the band structure, enabling carrier transfer into the central active layer while suppressing non-radiative leakage paths typically found at open surfaces\cite{reparaz2010recombination}.

\begin{figure}[ht!]
     \centering
     \includegraphics[width=0.95\textwidth]{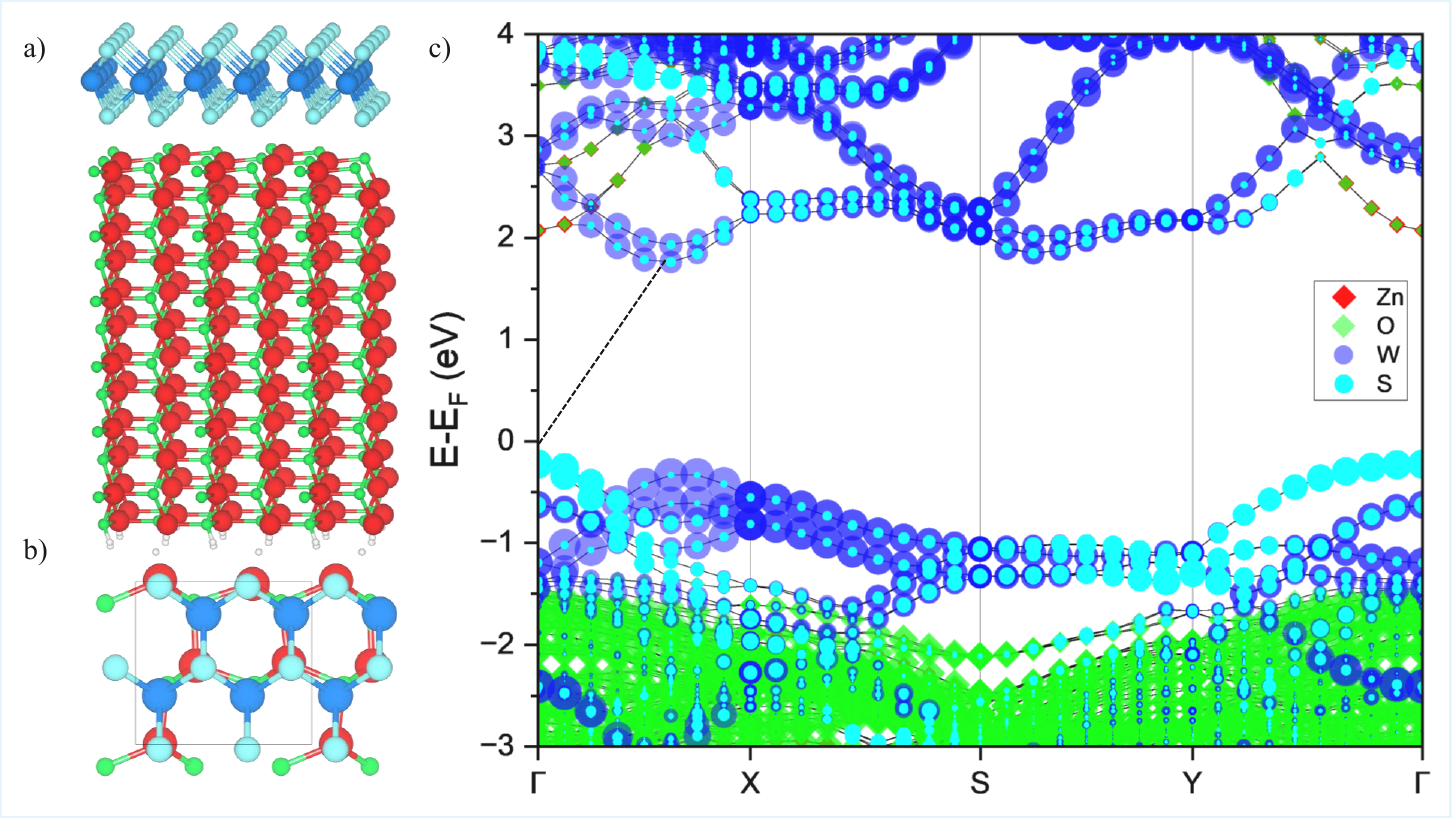}
     \caption{Structural and electronic characterization of the $WS_2/ZnO(11\bar{2}0)$ $a$-plane heterostructure. a) side-view supercell, b) top-view illustrating the full-eclipsed stacking, and c) element-resolved band structure. The configuration results in transition to indirect bandgap with the value of 2.09~eV.}
     \label{fig:a_plane}
\end{figure}

In contrast, our structural model of the $(11\bar{2}0)$ $a$-plane facet (see Fig.~\ref{fig:a_plane}) subjects the WS$_2$ layer to a severe lattice mismatch of 5.95\%. As seen in the top view of  Fig.~\ref{fig:a_plane}b), the two bonding networks of WS$_2$ and ZnO adopt a fully eclipsed conformation. While this stacking maximizes interfacial orbital overlap, the accompanying significant anisotropic strain ($110$~meV/\AA$^2$) induce substantial changes to the WS$_2$ band structure.
 
The structural constraints force an upward shift of the valence band near the $\Gamma$-point, driven by sulfur $p$-orbitals experiencing reduced orbital overlap in the $z$-direction. This results in a direct-to-indirect bandgap transition, with the effective gap narrowing to 2.09~eV. The direct optical band gap is 0.1X eV larger, and thus more comparable to the band gap encountered on the $m$-plane.

Regarding the Zn-terminated (0001) surface, in a gaseous environment, this surface can be stabilized by the adsorption of negative ions like $O^{2-}$ and $OH^-$\cite{dulub2003novel}, while under UHV conditions different kinds of surface reconstructions (deficient surface stoichiometries) occur\cite{rohrer_thermodynamic_2021}. Our $(0001)+OH$ surface model presents a significant departure from the non-polar models. The hydroxylated polar interface transitions to a staggered stacking orientation, resulting in a vertical offset between the W and Zn atomic positions. While this conformation typically represents a low-energy, stable configuration (interface energy of $-112$~meV/\AA$^2$ and the lowest lattice mismatch of only 2.04\%), the electronic landscape is dominated by the strong dipoles of the $-OH$ groups rather than the geometry.

\begin{figure}[ht!]
     \centering
     \includegraphics[width=0.95\textwidth]{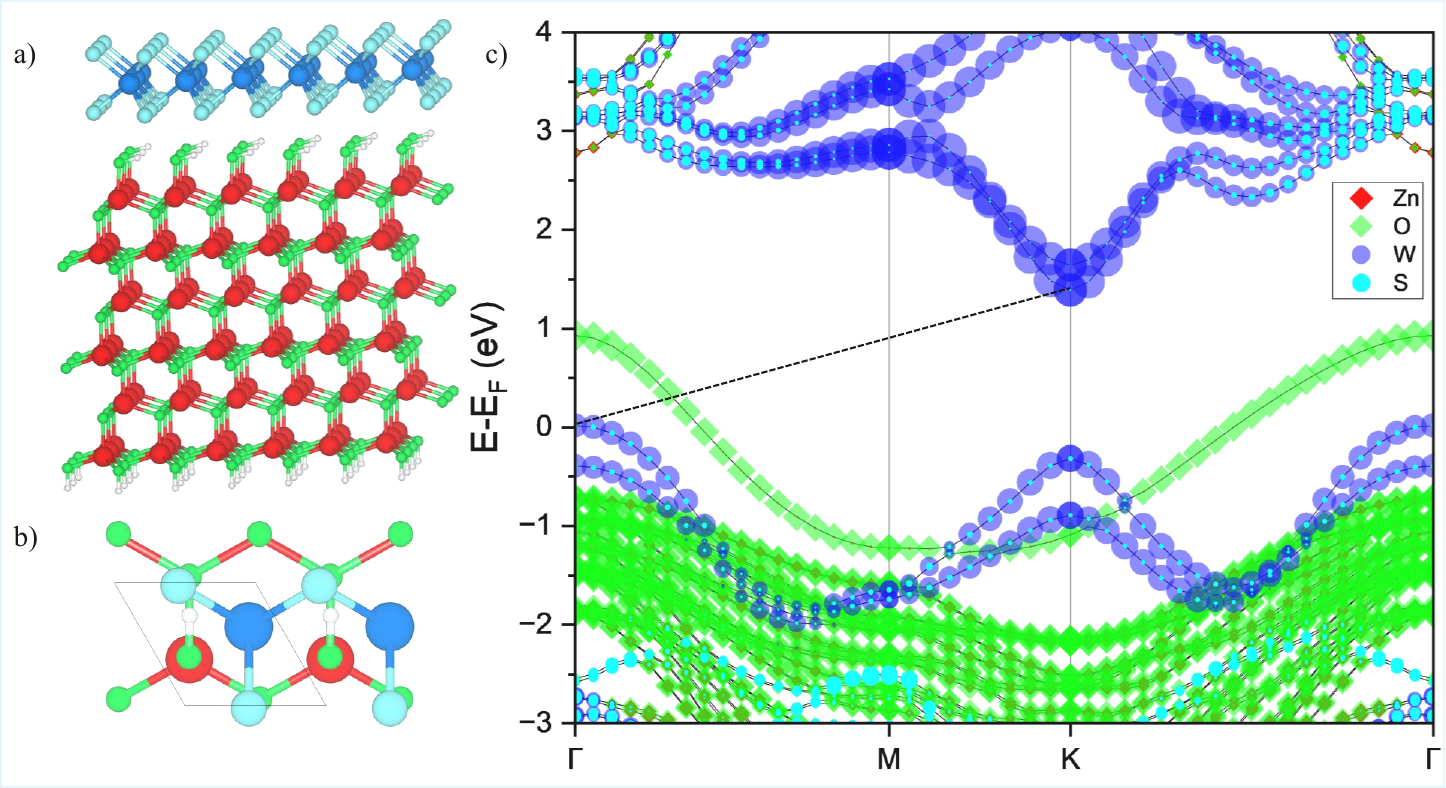}
     \caption{Structural and electronic characterization of the $WS_2/ZnO(0001)$ interface. a) side-view supercell, b) top-view illustrating the staggered stacking, and c) element-resolved band structure. Despite a stable staggered stacking, strong interfacial dipoles trigger a shift to a staggered Type-II alignment with a reduced indirect gap of 1.71~eV, indicated by the line.}
     \label{fig:polar}
\end{figure}

As seen from Fig.~\ref{fig:polar}, the WS$_2$ bandgap transforms to an indirect 1.71~eV, a significant reduction from the bulk 2.51~eV. This reduction is not strain-driven (as the strain energy is a minimal 5~meV/\AA$^2$) but is instead the result of strong interfacial dipoles originating from the $-OH$ passivation groups. These dipoles reconfigure the electrostatic landscape, inducing a shift from type-I to a staggered type-II alignment. 

Furthermore, the band structure reveals the emergence of an interfacial band originating from the oxygen $p$-orbitals of the hydroxylated surface near the Fermi level (see Fig.~\ref{fig:polar}c). This delocalized OH-band can act as a non-radiative leakage channel for holes, allowing them to bypass the WS$_2$ active layer\cite{reparaz2010recombination}. Consequently, while the polar $c$-plane offers superior structural stability, it is electronically unsuitable for LED applications due to the loss of carrier confinement and the transition to an indirect recombination regime.

\begin{table}[ht]
\centering
\small 
\caption{Calculated Structural and Electronic Properties of WS$_2$/ZnO Heterostructures across different crystallographic facets.}
\label{tbl:facets}
\begin{tabularx}{\textwidth}{lcccccX}
\toprule
\textbf{Model} & \textbf{Mismatch (\%)} & \textbf{Dist. (\AA)} & \textbf{$E_{int}$*} & \textbf{$E_{strain}$*} & \textbf{Nature} & \textbf{Alignment} \\ 
\midrule
$m$-plane $(10\bar{1}0)$ & 2.67 & 2.42 & $-43$ & 9 & Direct & Type-I \\
m-sandwich & 2.67 & 2.47 & - & 12 & Direct & Type-I \\
$a$-plane $(11\bar{2}0)$ & 5.95 & 2.64 & $-35$ & 110 & Indirect & Type-I \\
polar $(0001)$+OH & 2.04 & 2.46 & $-112$ & 5 & Indirect & Type-II \\
\bottomrule
\end{tabularx}
\begin{flushleft}
\footnotesize *Units for Interface Energy ($E_{int}$) and Strain Energy ($E_{strain}$) are meV/\AA$^2$.
\end{flushleft}
\end{table}

There is a potential trade-off where the choice of facet modulates the electronic band structure through strain and electrostatic potentials (band bending) rather than just structural fit. The choice of non-polar facet can act as a "tuning knob" for the electronic interface. $m$-plane ($10\bar{1}0$) induces an upward bending of the VBM, effectively narrowing the surface band gap (Fig. S3). While $a$-plane ($11\bar{2}0$) causes an upward bending of the CBM, opening the surface band gap.

Literature gives some additional hints to why a polar ZnO substrate may be unfavorable. While the formation of a full OH-terminated overlayer of (0001) polar facet can theoretically satisfy the electron counting rule, partial hydrogenation or metastable H-phases often lead to Fermi level pinning and surface conductivity. This metallization effectively turns the interface into a non-radiative zone, absorbing carriers that would otherwise contribute to light emission\cite{silva_hydrogen-induced_2018}. Moreover, while our DFT models provide a baseline for interface stability, the synthesis of ZnO introduces complexities in the real material. At elevated temperatures typical for growth ($>300^\circ C$), the Zn-polar surface can form a disordered, Zn-deficient "quasi-liquid" layer\cite{wang_stability_2021}. This suggests that growing 2D materials on polar ZnO at high temperatures involves depositing onto a highly dynamic, disordered surface rather than a rigid crystalline template.

\subsection{Thermodynamic Stability and Electronic Properties of Interfacial Point Defects}

The $(10\bar{1}0)$ $m$-plane facet was selected as the substrate for our defect investigation because it serves as the most electronically stable template identified in our initial screening. Unlike the polar $(0001)$ surface, which requires hydroxyl layers to compensate for surface charge, the $m$-plane is autocompensated and non-polar. This lack of polarity is crucial as it provides a clean background that allows us to isolate the specific impact of defects like $V_{S}$ or $H_{i}$ without the interference of the interfacial bands or potential shifts seen on the polar facets. Finally, since the $m$-plane preserves the direct bandgap and type-I alignment necessary for light emission, performing the defect study on this facet provides a realistic baseline for how native and impurity-driven states limit the performance of an optimized LED structure.

While the pristine $(10\bar{1}0)$ $m$-plane heterostructure provides an ideal type-I alignment, the optoelectronic performance of real-world devices is fundamentally dictated by the presence of native and impurity-driven point defects. To investigate these effects, we utilized the 'sandwiched' ZnO/WS$_2$/ZnO model (Fig. S4). From a computational perspective, the elimination of the vacuum region simplifies the application of periodic defect corrections and effectively mimics the high-dielectric environment of an encapsulated 2D layer. Physically, this symmetric environment provides uniform environmental dielectric screening, which stabilizes the quasiparticle renormalization of the WS$_2$ layer.

\begin{figure}[!ht]
    \centering
    \includegraphics[width=0.6\textwidth]{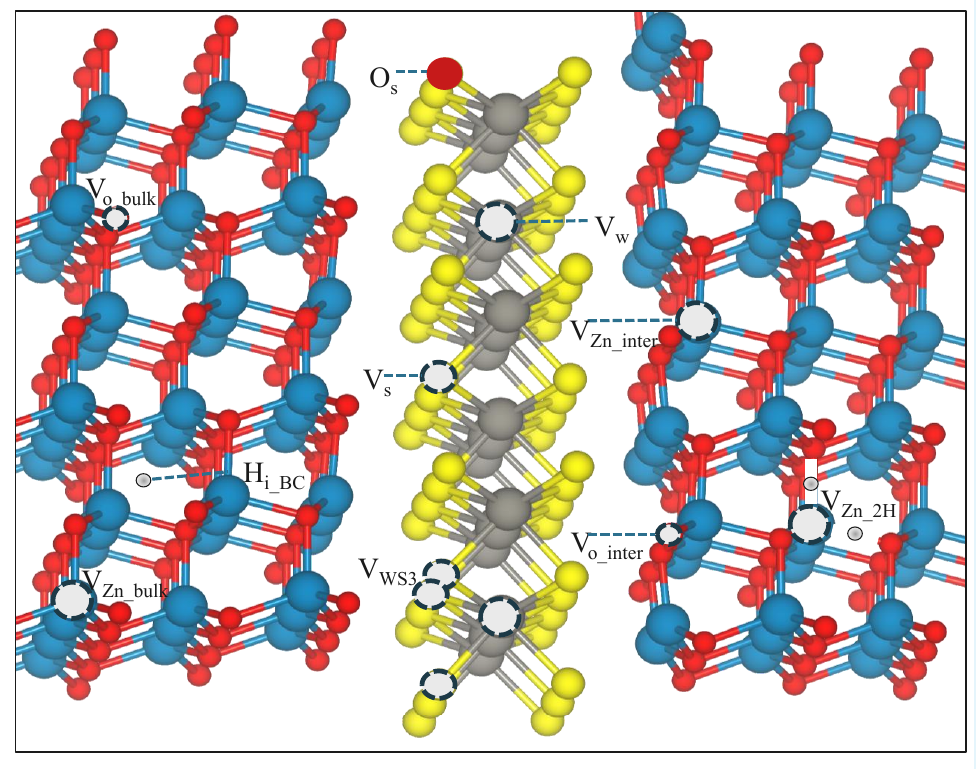}
    \caption{Schematic of analyzed point defects in WS$_2$/ZnO heterostructure.}
    \label{fig:schematic_defect}
\end{figure}

We performed a thermodynamic and electronic investigation of ten distinct defect species across the WS$_2$/ZnO junction (Fig.~\ref{fig:schematic_defect}), categorized into two groups: For WS$_2$, we consider the intrinisc sulfur vacancy (V$_S$) which has the lowest formation energy of point defects in WS$_2$, the tungsten vacancy (V$_W$) and a multi-atomic vacancy cluster (V$_{WS3}$). Moreover, oxygen-on-sulfur substitution (O$_S$) seems likely due to the presence of ZnO. For the defects in ZnO itself, our selection is guided by the literature on point defects: It is known that the oxygen vacancy (V$_O$) has a low formation energy over a wide range of conditions. Using DFT calculations, it was found to induce a charge transfer level in the upper part of the ZnO gap, yet too far away from the CBM to be responsible for the frequently encountered n-type conductivity of ZnO\cite{janotti_fundamentals_2009}. Moreover, it was shown that the zinc vacancy (V$_{Zn}$) becomes stable in n-type ZnO under oxygen-rich conditions. For this reason is was included in the analysis. For the heterostructures we aim to study, one needs to distinguish between vacancies in the bulk of ZnO and at the interface with WS$_2$. Since it is established that the n-type conductivity of native ZnO is in most cases due to atomic hydrogen impurities\cite{hofmann2002hydrogen}, we include the hydrogen interstitial at the bond center and the hydrogen-passivated zinc vacancy into this study. 

In the following sections, we evaluate the formation energies and the resulting electronic signatures of these species. We want to find out if and how the properties of defects in one material are affected by the presence of the other material, and in which way pairs of defects in both materials interact. By establishing the thermodynamic driving forces for these defects, we identify how the prevalence of localized in-gap states disrupts the pristine type-I confinement, introducing non-radiative recombination centers that compromise LED internal quantum efficiency.

\subsubsection{Energetics and Structural Relaxation}

To quantify the relative stability of various point defects and clusters within the sandwiched WS$_2$/ZnO heterostructure, we calculate the defect formation energy, $E_f(X^q)$, as a function of the chemical potentials ($\mu_i$) and the Fermi level ($E_F$):

\begin{equation}
E_f(X^q) = E_{tot}(X^q) - E_{tot}(\text{pristine}) - \sum_i n_i\mu_i + q(E_{VBM} + E_F) + \Delta_{corr}
\end{equation}
where $E_{tot}(X^q)$ and $E_{tot}(\text{pristine})$ are the total energies of the defect-containing and perfect supercells, respectively, and $n_i$ represents the number of atoms of species $i$ added or removed. In case of charged defects, a correction term, $\Delta_{corr}$, is applied using the Freysoldt, Neugebauer, and Van de Walle (FNV)\cite{freysoldt_first-principles_2014,freysoldt_first-principles_2018} method to account for periodic image interactions.

The chemical potentials are constrained by the stability of the constituent bulk phases, where $\Delta\mu_W + 2\Delta\mu_S = \Delta H_f(WS_2)$ and $\Delta\mu_{Zn} + \Delta\mu_O = \Delta H_f(ZnO)$. We evaluate two distincts growth conditions: (i) Anion-Poor (Metal-Rich), where $\Delta\mu_W=0$ and $\Delta\mu_{Zn}=0$; and (ii) Anion-Rich (Metal-Poor), where $\Delta\mu_S=0$ and $\Delta\mu_O=0$. Additionally, the formation of secondary competing phases such as $ZnS$, $WO_3$, and $ZnWO_4$ is considered to define additional thermodynamic limits for $\Delta\mu_W$, $\Delta\mu_{O}$  and $\Delta\mu_{S}$ (see Supplementary Information).

\begin{figure}[ht!]
    \centering
    \includegraphics[width=0.6\textwidth]{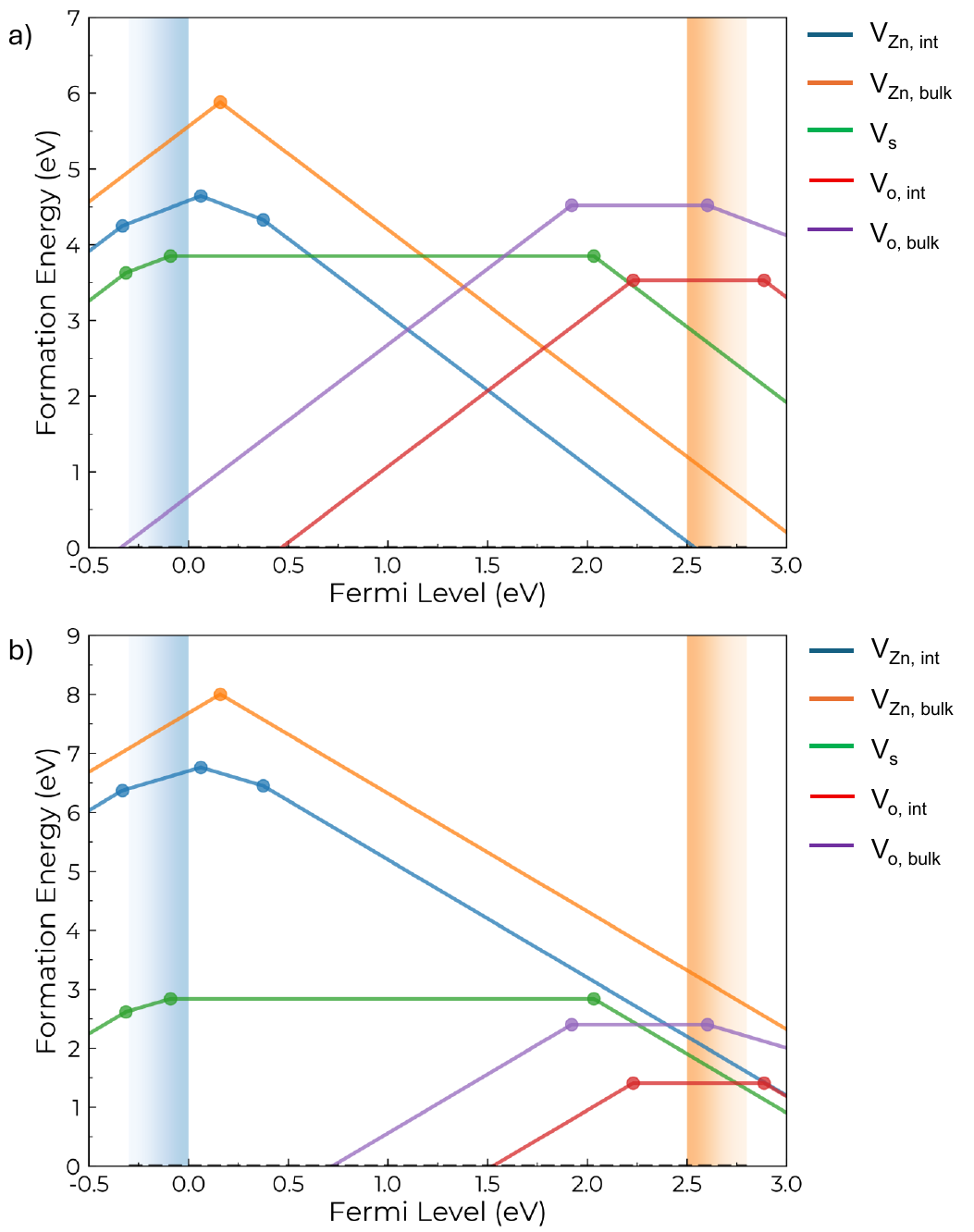}
    \caption{Defect formation energies as a function of Fermi level for the $\text{WS}_2/\text{ZnO}$ heterostructure under (a) anion-rich and (b) anion-poor growth conditions. The Fermi level is referenced to the heterostructure valence band maximum (VBM), with the shaded region indicating the conduction band.}
    \label{fig:charged_defect}
\end{figure}

Our first-principles calculations reveal that the WS$_2$/ZnO junction is a chemically active region, with formation energies for most point defects significantly lower at the interface than in the bulk transport layers. The energetic landscape for all ten defect species is summarized in Table~\ref{tab:comprehensive_defects}, while selected point defects are displayed as function of the Fermi level position in Fig.~\ref{fig:charged_defect}. 

The sulfur ($\text{V}_{\text{S}}$) and oxygen ($\text{V}_{\text{O}}$) vacancies represent the energetically accessible native defects within the heterostructure. We start by discussing the neutral states of these vacancies. Depending on the chemical potentials, the energy required to form $\text{V}_{\text{S}}$ ($\text{V}_{\text{O}}$) lies between 2.84 and 3.85~eV, and between 1.41 and 3.53~eV, respectively, see Table~\ref{tab:comprehensive_defects}. This falls within a similar range compared to literature values for WS$_2$~\cite{haldar_systematic_2015}. Notably, $\text{V}_{\text{O}}$ and zinc vacancies ($\text{V}_{\text{Zn}}$) are energetically easier to form at the interface compared to the bulk, suggesting the junction acts as a preferential site for defect accumulation. In the case of $\text{V}_{\text{Zn}}$, the formation energy is lower at the interface across both anion-rich and anion-poor regimes. 

\begin{table}[!ht]
\centering
\small 
\caption{Calculated defect formation energies ($E_{\text{f}}$), nearest-neighbor bond distances ($d_{\text{NN}}$), and percentage deviations ($\Delta d_{\text{NN}}$) for native and passivated defect sites.}
\label{tab:comprehensive_defects}
\begin{tabular}{llcccc}
\toprule
\multirow{2}{*}{\textbf{Defect Type}} & \textbf{Anion-Rich}                 & \textbf{Anion-Poor}                 & \multicolumn{3}{c}{\textbf{Structural Relaxation}} \\ 
\cmidrule(l){4-6}
                     & \textbf{$E_{\text{f}}$ (eV)}        & \textbf{$E_{\text{f}}$ (eV)}        & \textbf{Neighbor Atom} & \textbf{Distance (\AA)} & \textbf{$\Delta d_{\text{NN}}$ (\%)} \\ 
\midrule
                            &         &         & $\text{V}_{\text{S}}\text{--W}_{1}$  & 2.31 & $-4.6$  \\
V$_{\text{S}}$              & 3.85    & 2.84    & $\text{V}_{\text{S}}\text{--W}_{2}$  & 2.45 & $+1.1$  \\
                            &         &         & $\text{V}_{\text{S}}\text{--W}_{3}$  & 2.44 & $+0.9$  \\ \midrule
                            &         &         & $\text{V}_{\text{O}}\text{--Zn}_{1}$ & 1.57 & $-16.8$ \\
V$_{\text{O, int}}$         & 3.53    & 1.41    & $\text{V}_{\text{O}}\text{--Zn}_{2}$ & 1.67 & $-13.3$ \\
                            &         &         & $\text{V}_{\text{O}}\text{--Zn}_{3}$ & 1.67 & $-13.6$ \\ \midrule
                            &         &         & $\text{V}_{\text{O}}\text{--Zn}_{1}$ & 1.70 & $-14.6$ \\
V$_{\text{O, bulk}}$        & 4.52    & 2.40    & $\text{V}_{\text{O}}\text{--Zn}_{2}$ & 1.72 & $-13.4$ \\
                            &         &         & $\text{V}_{\text{O}}\text{--Zn}_{3}$ & 1.72 & $-13.6$ \\ 
                            &         &         & $\text{V}_{\text{O}}\text{--Zn}_{4}$ & 1.75 & $-12.1$ \\ \midrule
                            &         &         & $\text{V}_{\text{Zn}}\text{--O}_{1}$ & 1.97 & $+4.2$  \\
V$_{\text{Zn, int}}$        & 4.87    & 6.99    & $\text{V}_{\text{Zn}}\text{--O}_{2}$ & 1.98 & $+2.4$  \\
                            &         &         & $\text{V}_{\text{Zn}}\text{--O}_{3}$ & 1.98 & $+2.6$  \\ \midrule
                            &         &         & $\text{V}_{\text{Zn}}\text{--O}_{1}$ & 2.14 & $+7.1$  \\
V$_{\text{Zn, bulk}}$       & 6.05    & 8.18    & $\text{V}_{\text{Zn}}\text{--O}_{2}$ & 2.14 & $+7.2$  \\
                            &         &         & $\text{V}_{\text{Zn}}\text{--O}_{3}$ & 2.16 & $+8.3$  \\ 
                            &         &         & $\text{V}_{\text{Zn}}\text{--O}_{4}$ & 2.14 & $+7.2$  \\ \midrule
                            &         &         & $\text{O--W}_{1}$                    & 2.05 & $-14.1$ \\
O$_{\text{S}}$              & $-0.53$ & 0.58    & $\text{O--W}_{2}$                    & 2.09 & $-13.5$ \\
                            &         &         & $\text{O--W}_{3}$                    & 2.09 & $-12.9$ \\ \midrule
                            &         &         & $\text{V}_{\text{Zn}}\text{--O}_{1}$ & 1.80 & $-9.4$  \\
V$_{\text{Zn}}$ + 2H        & 0.96    & $-1.83$ & $\text{V}_{\text{Zn}}\text{--O}_{2}$ & 1.95 & $-2.0$  \\
                            &         &         & $\text{V}_{\text{Zn}}\text{--O}_{3}$ & 2.01 & $+0.9$  \\
                            &         &         & $\text{V}_{\text{Zn}}\text{--O}_{4}$ & 1.94 & $-2.5$  \\ \cmidrule{1-3}
H$_{\text{i}}$$^{a}$         & 2.56    & 2.56    &                                      &      &          \\ \cmidrule{1-3}
V$_{\text{W}}$$^{b}$         & 7.34    & 9.37    &                                      &      &          \\ \cmidrule{1-3}
V$_{\text{WS}_3}$$^{b}$      & 14.78   & 13.76   &                                      &      &          \\
\bottomrule
\end{tabular}
\begin{flushleft}
\footnotesize Variations are referenced against specific pristine baselines: 2.42~\AA\ for $\text{W--S}$ bonds; 1.89~\AA\ and 1.93~\AA\ for surface $\text{Zn--O}$ bonds; and 1.99~\AA\ for bulk tetrahedral $\text{Zn--O}$ bonds.\\
$^{a}$ Local geometric parameters are omitted for interstitial H$_{\text{i}}$ due to the lack of a fixed lattice site baseline comparable to substitutional configurations.\\
$^{b}$ Due to the exceptionally high formation energies of V$_{\text{W}}$ and V$_{\text{WS}_3}$ complexes, these configurations are thermodynamically suppressed under equilibrium growth conditions. Hence, detailed local coordinate analysis was omitted.
\end{flushleft}
\end{table}

The role of oxygen and hydrogen in modulating interfacial quality is significant. Given its low formation energy under anion-rich conditions ($-0.53$~eV), oxygen is predicted to naturally replace sulfur at the interface ($\text{O}_{\text{S}}$), likely leading to the formation of a $\text{WS}_{2-x}\text{O}_x$ transition layer~\cite{kotsakidis2019oxidation}. Furthermore, hydrogen impurities demonstrate a passivating effect on cation defects~\cite{kang2016light}; the $\text{V}_{\text{Zn}}\text{--}2\text{H}$ complex possesses a lower formation energy (ranging from $-1.83$~eV under anion-poor to 0.96~eV under anion-rich conditions) than the isolated $\text{V}_{\text{Zn}}$.

The bond-centered hydrogen interstitial ($\text{H}_{i\text{-BC}}$) exhibits a moderate formation energy of 2.56~eV, which is independent of the anion-rich or anion-poor conditions. As a known shallow donor in $\text{ZnO}$, the presence of $\text{H}_{i\text{-BC}}$ near the junction is expected to facilitate n-type conductivity and assist in electron injection into the $\text{WS}_2$ active layer~\cite{van2000hydrogen,janotti_hydrogen_2007}. Notably, while $\text{H}_{i\text{-BC}}$ is more stable than high-energy structural defects, it is less favorable than the $\text{V}_{\text{Zn}}\text{--}2\text{H}$ complex under anion-poor conditions ($-1.83$~eV), highlighting the dual role of hydrogen as both a dopant and a passivating agent in these heterostructures.

Consistent with their larger structural disruption, the tungsten vacancy ($\text{V}_{\text{W}}$) and the $\text{V}_{\text{WS}_3}$ cluster exhibit the highest formation energies ($>7.3$~eV), indicating they are less likely to form under equilibrium growth conditions compared to other point vacancies.

The calculated structural relaxations presented in Table~\ref{tab:comprehensive_defects} also show how the local environment shifts around each defect site. In the WS$_2$ layer, the sulfur vacancy (V$_S$) provides direct evidence of the substrate's influence. In a free-standing WS$_2$ monolayer, the relaxation is symmetric, with all three tungsten neighbors pulling inward by roughly $0.9\%$. At the interface, however, this symmetry is broken. While the average bond change remains $-0.87\%$, the $\text{V}_{\text{S}}\text{--W}_{1}$ distance undergoes a strong lateral contraction of $-4.57\%$, while $\text{V}_{\text{S}}\text{--W}_{2}$ and $\text{V}_{\text{S}}\text{--W}_{3}$ shift outward by about $+1.0\%$. This asymmetric relaxation indicates that the rigid vertical alignment of the ZnO $m$-plane limits vertical movement, forcing the lattice strain to adjust horizontally along the surface channels. When oxygen replaces sulfur at this site ($O_{\text{S}}$), the tungsten framework itself remains stable, with the metal atoms shifting by only $2\%$ to $3\%$ ($\sim0.10$~\AA) from their original positions. Instead, the shorter bond lengths observed—$2.05$~\AA\ for $\text{O--W}_{1}$ and $2.09$~\AA\ for $\text{O--W}_{2,3}$—are simply due to the fact that an oxygen atom is physically smaller than a sulfur atom. This localized uniform shift effectively preserves the original mirror symmetry of the site.

Within the substrate, the relaxation depends directly on whether an anion or a cation is missing~\cite{janotti_fundamentals_2009, marrocchelli2012understanding, lany2005anion}, mapping explicitly across the tetrahedral coordination environments in the bulk phases. Removing an oxygen atom ($\text{V}_{\text{O}}$) causes the neighboring zinc ions to contract inward toward the empty site. This collapse is most pronounced at the interface, where the $\text{V}_{\text{O}}\text{--Zn}_{1}$ distance contracts by $-16.83\%$, compared to roughly $-13.5\%$ for $\text{V}_{\text{O}}\text{--Zn}_{2}$ and $\text{V}_{\text{O}}\text{--Zn}_{3}$. This localized structural relaxation directly explains the 1.0~eV drop in formation energy observed for $\text{V}_{\text{O, int}}$ (1.41~eV) relative to $\text{V}_{\text{O, bulk}}$ (2.40~eV), demonstrating that the structural flexibility of the interface drives defect stabilization. At the interface, surface atoms have fewer neighbors and more freedom to move. In the bulk, the surrounding 3D crystal lattice adds mechanical resistance, keeping the contractions of all four tetrahedral neighbors more uniform, ranging between $-12.06\%$ ($\text{V}_{\text{O}}\text{--Zn}_{4}$) and $-14.61\%$ ($\text{V}_{\text{O}}\text{--Zn}_{1}$).

Conversely, the oxygen atoms adjacent $V_{\text{Zn}}$ shift outward. This expansion is rather uniform across the four coordinating neighbors of the bulk phase. All four neighbors ($\text{V}_{\text{Zn}}\text{--O}_{1,2,3,4}$) exhibit tight, consistent outward displacements ranging between $+7.08\%$ and $+8.33\%$ ($2.14$~\AA\ to $2.16$~\AA). In contrast, the interface restricts this movement, allowing a lower-amplitude expansion across its remaining truncated coordination shell ($+2.36\%$ to $+4.23\%$).

In the bulk zinc vacancy complexes, introducing hydrogen modifies these strain fields. For the bulk zinc vacancy passivated by two hydrogen atoms ($V_{\text{Zn}}\text{-}2\text{H}$), the two protons form hydroxyl ($\text{O--H}$) bonds directly with the $\text{V}_{\text{Zn}}\text{--O}_{2}$ and $\text{V}_{\text{Zn}}\text{--O}_{3}$ neighbors. This localized chemical passivation neutralizes a portion of the local charge, leaving $\text{O}_{2}$ close to its pristine parameter with a small contraction of $-2.01\%$, while geometric constraints balance the position of $\text{O}_{3}$, resulting in a minor expansion of $+0.92\%$. The remaining two diagonal oxygen neighbors ($\text{V}_{\text{Zn}}\text{--O}_{1}$ and $\text{V}_{\text{Zn}}\text{--O}_{4}$) do not bind to hydrogen; instead, they accommodate the asymmetric strain fields of the reconstructed site. The unbonded neighbor $\text{V}_{\text{Zn}}\text{--O}_{4}$ undergoes a small contraction of $-2.52\%$, while the counterpart diagonal neighbor $\text{V}_{\text{Zn}}\text{--O}_{1}$ absorbs the bulk of the relaxation, contracting inward by $-9.42\%$ ($1.80$~\AA).

By contrast, bond-centered interstitial hydrogen (H$_{i\text{-BC}}$) inserts directly into a Zn--O bond rather than sitting at a vacant site. This bond insertion breaks the original Zn--O bond, increasing the atomic separation by 48.24\% to 2.95~\AA. The hydrogen binds directly to the oxygen atom to form an O--H bond, which pushes the decoupled zinc atom away from its ideal lattice position.

These structural trends demonstrate that while native vacancy geometries are governed by the local electrostatic environment of the heterostructure interface, hydrogen interactions are determined by distinct chemical pathways. Specifically, hydrogen can either mitigate lattice strain through local charge passivation or significantly disrupt the local oxide coordination via interstitial insertion.

\subsubsection{Electronic Signatures and Charge Transition Levels}

\begin{figure}[!ht]
    \centering
    \includegraphics[width=0.6\textwidth]{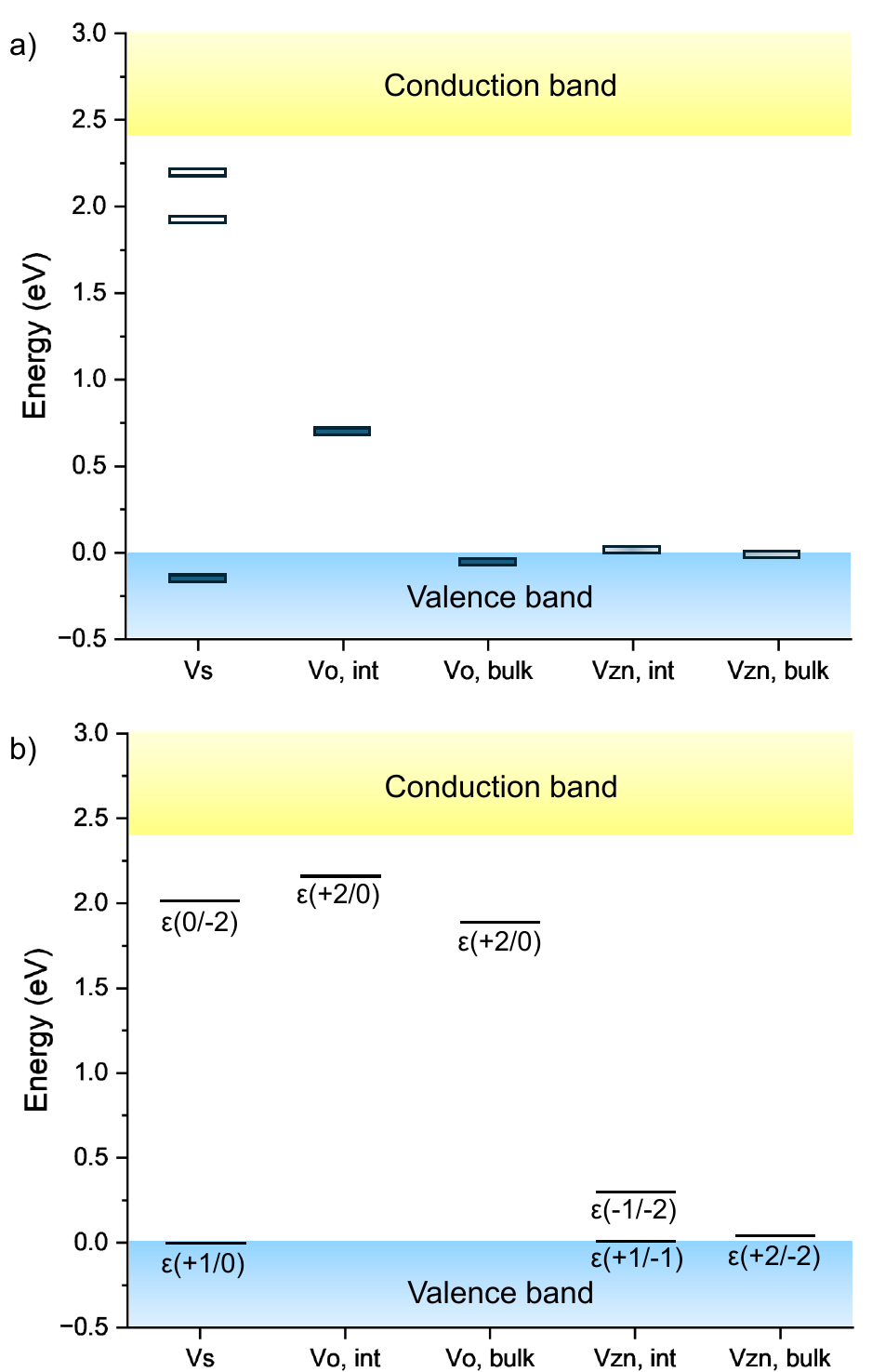}
    \caption{Electronic and thermodynamic characterization of single point defects in the WS$_2$/ZnO heterostructure. a) Single-particle generalized Kohn-Sham defect levels for selected vacancy species. Filled bars indicate fully occupied in-gap states, while open bars represent unoccupied trapping levels. The gradient-shaded bars for $V_{Zn\_inter}$ and $V_{Zn\_bulk}$ denote partially occupied states situated near the valence band maximum (VBM). b) Calculated thermodynamic charge transition levels (CTLs), $\epsilon(q/q')$, for corresponding interfacial and bulk defects.}
    \label{fig:defect_level}
\end{figure}

A primary objective of this study is to determine how the electronic signatures of native defects evolve when transitioning from isolated bulk or monolayer environments to an integrated WS$_2$/ZnO heterojunction. The calculated defect levels shown in Fig.~\ref{fig:defect_level}a) show moderate deviations from literature results of isolated materials, suggesting that the optoelectronic quality of the WS$_2$ emitter is affected by the stoichiometric condition of the  ZnO substrate. % (Fig.~\ref{fig:defect_level}).

In the generalized Kohn-Sham (gKS) method used to calculated electronic states, the sulfur vacancy (V$_S$) introduces an occupied level buried at $-0.15$~eV and a pair of deep unoccupied states at $1.92$~eV and $2.20$~eV. The presence of these states is in reasonable agreement with previous experimental findings for WS$_2$ of different substrates \cite{schuler_large_2019} as well as theoretical results for free-standing WS$_2$ using different density functionals \cite{haldar_systematic_2015}.

The prevalence of $V_{O}$ and $V_{S}$ at the interface is particularly significant because their electronic signatures undergo substantial change due to the local environment of the heterostructure. A central finding of this study is the shift of electronic gKS levels associated with the oxygen vacancy ($V_{O}$). In the $\text{ZnO}$ layer away from the junction ($V_{\text{O, bulk}}$), the vacancy introduces an occupied state situated just below the VBM of $\text{WS}_2$ at $-0.05$~eV (corresponding to roughly 0.65~eV above the VBM of $\text{ZnO}$). However, as the vacancy is positioned at the interface ($V_{O\text{-inter}}$), the more shallow potential at the interface lifts this level state into the $WS_{2}$ bandgap, appearing as an occupied mid-gap level at $0.7$~eV (see Fig.~\ref{fig:defect_level}a).

These deep-level states act as the primary bottlenecks for internal quantum efficiency by enabling Shockley-Read-Hall (SRH) recombination (see Ref.~\cite{kangsabanik_defect-assisted_2026} for information how to calculate recombination rates). Because $V_{O\text{-inter}}$ is both deep ($0.7$~eV) and occupied, it functions as a potential hole trap. Its energy separation from the VBM is large enough to suppress thermal re-emission, favoring non-radiative decay through multi-phonon emission. Simultaneously, the deep unoccupied states of $V_{S}$ ($1.9$--$2.2$~eV) act as persistent electron traps\cite{bretscher2021rational, daniel2025mechanism} and can give rise to photoluminescence from trapped states \cite{schuler_electrically_2020}. The synergy between these interfacial donor and acceptor traps establishes high-density non-radiative pathways that effectively suppress radiative excitonic recombination in the active WS$_2$ layer.

As seen from the right of Fig.~\ref{fig:defect_level}a), zinc vacancies ($V_{\text{Zn}}$) in the heterostructure exhibit a different behavior, acting as shallow acceptors. The bulk vacancy ($V_{\text{Zn, bulk}}$) introduces a partially occupied state just below the VBM at $-0.02$~eV, whereas the interfacial counterpart ($V_{\text{Zn, int}}$) shifts slightly above the band edge to $0.02$~eV. This partial occupancy represents a key feature of these cation defects that becomes important for the electronic mechanism involving H$_i$.

A comparative analysis of the electronic defect levels and the thermodynamic charge transition levels (CTLs) reveals the complex nature of carrier trapping in the WS$_2$/ZnO heterostructure. While the band structure, in particular if calculated in a generalized Kohn-Sham (gKS) scheme with a partially self-interaction-free hybrid functional, as is the case here, can be compared to spectroscopic data, the CTLs represent the specific Fermi level positions where a defect transitions between different charge states (Fig.~\ref{fig:defect_level}b). They define the thermodynamic boundaries that determine which ionization state is most stable under equilibrium conditions.

Looking at the thermodynamic CTLs in Fig.~\ref{fig:defect_level}b, we see a clear departure from bulk oxide behavior. In a bulk ZnO crystal, V$_{Zn}$ typically goes through separate single-electron transitions ($\epsilon(0/-)$ and $\epsilon(-/2-)$) in the lower half of the band gap~\cite{janotti_fundamentals_2009}. Here, however, $V_{\text{Zn, bulk}}$ stays pinned in its fully ionized $-2$ charge state across our entire scanned window. One has to keep in mind that the VBM of the heterostructure, chosen as the zero of the energy scale, is in fact the VBM of $\text{WS}_2$ while the  lower-lying VBM of ZnO is outside the energy range displayed. 

The interface vacancy ($V_{\text{Zn, int}}$) tells a different story, with its transition levels reaching into the band gap. This shallow position relative to the heterostructure VBM closely mirrors what is seen in bulk $\text{ZnO}$ calculations~\cite{janotti_fundamentals_2009}. Because $V_{\text{Zn, int}}$ is in direct contact with the $\text{WS}_2$ monolayer, its defect wavefunctions strongly hybridize with those higher-lying $\text{WS}_2$ valence states. This orbital mixing essentially hooks the interface defect transitions to the $\text{WS}_2$ valence band edge. 

This alignment brings the single-particle gKS levels and thermodynamic CTLs into good agreement for the interface defect. Since the partially occupied state at $0.02$~eV sits right on the edge of the VBM, it allows for charge transfer with the valence band, matching the shallow thermodynamic thresholds perfectly. Ultimately, these results show that $V_{\text{Zn}}$ operates as an active shallow acceptor near the band edge, explaining the background p-type tendencies or partial compensation of electron doping often encountered at these oxide boundaries. 

For the sulfur vacancy (V$_S$), the neutral ($q=0$) charge state is stable across almost the entire band gap. Its thermodynamic transition level, $\epsilon(0/-2)$, sits high up at 1.93~eV, meaning the defect stays neutral from the valence band edge through more than 80\% of the gap. Interestingly, this direct jump to the $-2$ state is quite different from what happens in an isolated $\text{WS}_2$ monolayer, where the vacancy typically charges up in separate single-electron steps through an intermediate $-1$ state~\cite{lin2018revealing, khalid2024deep}. This change can be explaining by the surrounding dielectric environment. In a standalone monolayer, the lack of out-of-plane screening creates a Coulomb penalty ($U$) when there is an addition of extra electron into the defect site. When we put the monolayer on a high-dielectric $\text{ZnO}$ substrate, however, the extra screening will weaken this electron-electron repulsion. Once this Coulomb energy drops below the energy gained by local structural changes, the intermediate $-1$ state can become unstable, going directly into the $-2$ state. However, even with this change in charging mechanics, the transition stays high in the gap -- a direct consequence of the structural rigidity typical of 2D transition metal dichalcogenides~\cite{komsa2015native, khalid2024deep}. Because the tungsten network is stiff, the lattice relaxation energy is too small to drag the level deep into the mid-gap, keeping it well-aligned with the deep electron-trapping features reported by Lin \textit{et al.}~\cite{lin2018revealing}.

Similarly, the interfacial $V_{O\_inter}$ exhibits an occupied gKS level at $0.7$~eV, yet its primary thermodynamic transition $\epsilon(+2/0)$ occurs much higher at $2.23$~eV. Differences of similar size ($\sim 1$ eV) between single-particle levels and CTLs of V$_O$ have been reported previously \cite{gallino2010}. While the thermodynamic $\epsilon(+2/0)$ transition for $V_{\text{O, bulk}}$ sits lower at 1.92~eV. This large difference suggests that the $+2$ interfacial oxygen vacancy undergoes a substantial lattice distortion (57.2 \% outward relaxation) compared to its neutral state (16.8 \% inward relaxation). Thermodynamically, this implies that $V_{O\_inter}$ remains neutral only in n-type samples but looses two electrons already when the Fermi level drops to 0.5~eV below the CBM of WS$_2$, i.e. 1.2~eV below the CBM of ZnO. This observation matches with DFT studies of bulk ZnO, where the CTL of V$_O$ was located at 1.2 eV below the CBM of ZnO \cite{oba_defect_2008}. 

The energetic proximity of the $\epsilon(0/-2)$ transition of $V_{S}$ ($1.93$~eV) and $\epsilon(+2/0)$ transition of $V_{O}$ ($1.82$--$2.13$~eV) within the upper region of the bandgap create a high density of states for SRH recombination. Because these CTLs are deep, they act as quasi-permanent traps for carriers. In contrast, the near-VBM transitions of V$_{Zn}$ ensure that these defects contribute to mobile carrier density rather than non-radiative loss, identifying V$_{Zn}$ as a 'benign' defect compared to the $V_{S}$ and $V_{O}$ species.

Mitigating the detrimental effect of these interfacial traps is essential for high-efficiency LED operation. We find that the oxygen-on-sulfur (O$_S$) substitution is an effective passivation strategy. Driven by its low formation energy at the interface, oxygen naturally occupies sulfur vacancies, leading to defect saturation. As shown in the band structure analysis (Fig. S11), this substitution eliminates the in-gap states characteristic of V$_S$, effectively restoring the pristine electronic structure of the WS$_2$ layer. Similarly, the formation of $V_{Zn}\text{--}2H$ complexes successfully passivates the zinc vacancy, removing its deep-level signatures and clearing the gap for efficient carrier transport.

\begin{figure}[hb!]
    \centering
    \includegraphics[width=0.6\textwidth]{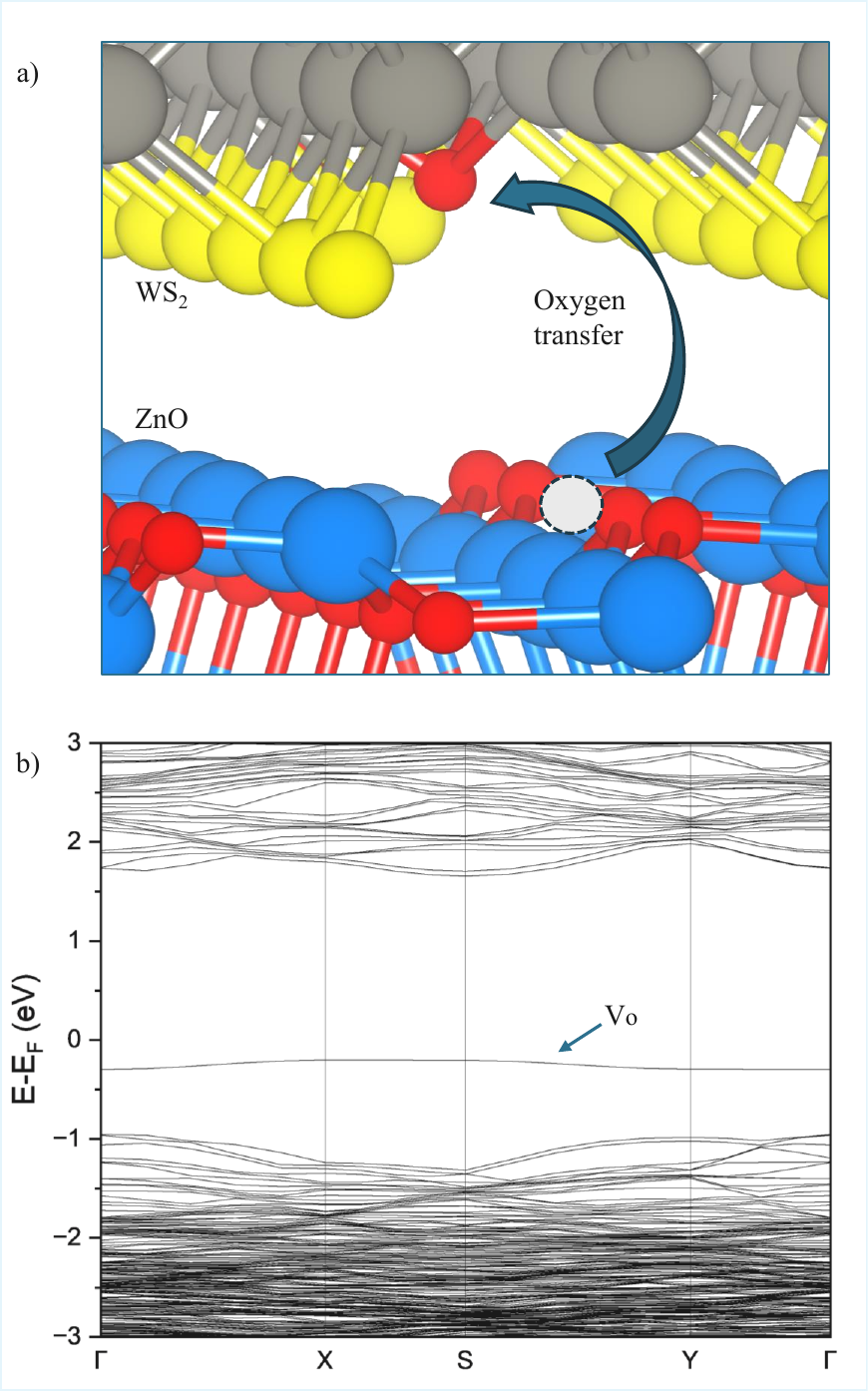}
    \caption{Interfacial oxygen transfer and electronic passivation mechanisms. a) Schematic representation of the oxygen transfer process from the ZnO substrate to the $WS_{2}$ monolayer, resulting in oxygen-on-sulfur ($O_{S}$) substitution. b) Calculated electronic band structure illustrating the defect state associated with the oxygen vacancy ($V_{O}$).}
    \label{fig:oxygen_migration}
\end{figure}

It is well-established that native zinc oxide exhibits inherent n-type conductivity, frequently attributed to unintentional shallow donor impurities such as bond-centered interstitial hydrogen ($H_{\text{i-BC}}$). Our electronic structure calculations for the $\text{WS}_2/\text{ZnO}$ heterostructure confirm this behavior, showing that the presence of $H_{\text{i-BC}}$ successfully shifts the system's Fermi level upward, pinning it close to the conduction band minimum (CBM) of the Type-I aligned junction.

In terms of the defect stability diagrams (Fig.~\ref{fig:charged_defect}), this reveals a charge compensation mechanism that depends directly on the growth environment. If $n$-type doping is attempted under anion-rich conditions, the low formation energy of $\mathrm{V}_{\mathrm{Zn}, \text{bulk}}''$ makes it the primary compensating acceptor that counteracts electron doping. This behavior shifts in the anion-poor regime. Despite their low formation energy across most of the bandgap, oxygen vacancies (V$_{O, bulk}$) cannot compensate $n$-type doping because they remain neutral ($q = 0$) for $E_{\text{F}}$ near the CBM. Instead, sulfur vacancies ($\mathrm{V}_{\mathrm{S}}''$) and bulk zinc vacancies ($\mathrm{V}_{\mathrm{Zn}, \text{bulk}}''$) act as the compensating species. Both defects exhibit steep negative slopes at high Fermi levels, reflecting their capacity to trap conduction-band electrons.

\subsubsection{Interfacial Oxygen Transfer and Self-Passivation}

Oxygen substitution on sulfur vacancy has been observed in scanning tunneling microscopy (STM) as a prominent point defect in WS$_2$\cite{barja2019identifying}, which can arise from chemisorption followed by dissociation of O$_2$ on V$_S$ site\cite{lu2015atomic, liu2015atomistic}. Given that the underlying ZnO substrate can serve as an immediate oxygen reservoir, we examined the thermodynamic feasibility of this interfacial redox reaction of oxygen atom occupation on pre-existing sulfur vacancies (V$_S$) at the adjacent WS$_2$ interface (Fig.\ref{fig:oxygen_migration}).  

Treating the V$_S$-defected heterostructure as the reference state, the transition to the $O_S + V_{O\text{-inter}}$ configuration is exothermic by $0.91$~eV. This energy gain suggests a strong thermodynamic drive for interfacial self-passivation, where ZnO suppresses sulfur-vacancy-related defects during thermal processing.

The electronic consequences of oxygen transfer are significant. Band structure analysis of the $O_S + V_{O\text{-inter}}$ pair model shows the complete elimination of the deep unoccupied V$_S$ traps originally found at 1.8 and 2.2~eV. While this process restores the conduction band integrity of the WS$_2$ layer, it leaves behind an occupied mid-gap state at $\approx0.7$~eV associated with the interfacial oxygen vacancy. These findings indicate that while oxygen transfer effectively neutralizes electron traps, the resulting $O$-deficiency at the ZnO surface remains a persistent hole-trapping bottleneck for radiative efficiency.

\subsection{Interaction of Point Defects}

From a general perspective, the thermodynamic interactions of point defects are governed by distinct physical mechanisms that influence stabilization. For bulk $\mathrm{ZnO}$, literature indicates that these interactions are thermodynamic in nature. For instance, zinc interstitials ($\mathrm{Zn}_i^{\bullet\bullet}$), which are typically unstable as shallow donors due to high formation energies, can be thermodynamically stabilized through the formation of defects with oxygen vacancies ($\mathrm{V}_{\mathrm{O}}$)\cite{kim2009rich}. These bound pairs---such as the zinc antisite ($\mathrm{Zn}_{\mathrm{O}}$), effectively a $(\mathrm{Zn}_i - \mathrm{V}_{\mathrm{O}})$ defect, or impurity-vacancy defects like $(\mathrm{Zn}_i - \mathrm{N})'$ (with a binding energy of $\sim 0.9$~eV)---prevent defect out-diffusion and significantly alter electrical compensation levels within the $\mathrm{ZnO}$ matrix.

This contrasts with defect interactions within the WS$_2$ monolayer, which are mediated by local strain fields and orbital hybridization. In a recent study\cite{frammolino_microscopic_2025}, the interplay of a finite concentration of O$_{\text{S}}$ with a sulfur vacancy was examined. It was found that the interaction between sulfur vacancies (V$_{\text{S}}$) and oxygen substituents (O$_{\text{S}}$), combined with modified $d\text{--}p$ orbital hybridization, pushes the V$_{\text{S}}$ defect resonance from the valence band deep into the bandgap. Unlike the fixed binding energies characteristic of ZnO defect pairs, these interactions result in fluctuating energy levels that depend on the specific proximity and density of surrounding substituents\cite{frammolino_microscopic_2025}.

\begin{table}[!ht]
\centering
\small
\caption{Calculated defect pair formation energies ($E_{\text{f}}$) and corresponding binding energies ($E_{\text{b}}$) for combined point defect configurations. Values for charged states ($q \neq 0$) are evaluated at a Fermi level of $E_{\text{F}} = 2.3$~eV relative to the valence band maximum.}
\label{tab:defect_complexes}
\resizebox{\textwidth}{!}{%
\begin{tabular}{lcccc}
\toprule
\textbf{Defect Pair} & \textbf{Charge ($q$)} & \textbf{Anion-Rich $E_{\text{f}}$ (eV)} & \textbf{Anion-Poor $E_{\text{f}}$ (eV)} & \textbf{Binding Energy $E_{\text{b}}$ (eV)} \\
\midrule
$(\mathrm{V}_{\mathrm{S}} - \mathrm{V}_{\mathrm{O,int}})$   & $0$  & 7.95  & 4.81  & $-0.58$ \\
$(\mathrm{V}_{\mathrm{S}} - \mathrm{V}_{\mathrm{O,bulk}})$  & $0$  & 8.93  & 5.79  & $-0.59$ \\
$(\mathrm{V}_{\mathrm{S}} - \mathrm{V}_{\mathrm{Zn,int}})$  & $0$  & 8.82  & 9.93  & $-1.03$ \\
$(\mathrm{V}_{\mathrm{S}} - \mathrm{V}_{\mathrm{Zn,bulk}})$ & $0$  & 10.44 & 11.55 & $-0.64$ \\
\midrule
$(\mathrm{V}_{\mathrm{S}} - \mathrm{V}_{\mathrm{O,int}})''$   & $-2$ & 9.48  & 6.34  & $-0.55$ \\
$(\mathrm{V}_{\mathrm{S}} - \mathrm{V}_{\mathrm{O,bulk}})''$  & $-2$ & 10.41 & 7.27  & $-0.61$ \\
$(\mathrm{V}_{\mathrm{S}} - \mathrm{V}_{\mathrm{Zn,int}})''''$ & $-4$ & 5.75  & 2.61  & $-0.22$ \\
$(\mathrm{V}_{\mathrm{S}} - \mathrm{V}_{\mathrm{Zn,bulk}})''''$& $-4$ & 6.67  & 3.53  & $-0.36$ \\
\bottomrule
\end{tabular}%
}
\end{table}

While individual formation energies provide a baseline for defect prevalence, the performance of WS$_2$/ZnO is ultimately determined by the collective behavior and interaction of these species at the junction. By calculating the binding energies of the most common pairings, specifically sulfur vacancies (V$_S$) with interfacial zinc (V$_{Zn}$) and oxygen (V$_O$) vacancies, we examine the physical drivers of defect clustering at the interface.

The binding energy for defect pairs is defined according to the following expression:

\begin{equation}
E_{\text{b}} = E_{\text{f}}\!\left( \text{Pair}_{AB}^q \right) - \left[ E_{\text{f}}\!\left( \text{Defect}_A^{q_1} \right) + E_{\text{f}}\!\left( \text{Defect}_B^{q_2} \right) \right]
\end{equation}
where $E_{\text{f}}(\text{Pair}_{AB}^q)$ is the formation energy of the combined defect pair in charge state $q = q_1 + q_2$, while $E_{\text{f}}(\text{Defect}_A^{q_1})$ and $E_{\text{f}}(\text{Defect}_B^{q_2})$ represent the formation energies of the isolated constituent defects.

Our calculations reveal consistently negative binding energies across all analyzed defect pairs, formally establishing the WS$_2$/ZnO interface as a definitive defect sink. As summarized in Table~\ref{tab:defect_complexes}, the interactions for the neutral baseline are consistently stronger at the interface compared to the bulk $\mathrm{ZnO}$ transport layers.

The data reveals that the $(\mathrm{V}_{\mathrm{S}} - \mathrm{V}_{\mathrm{Zn,int}})$ pair exhibits the strongest interaction with a binding energy of $-1.03$~eV. This high stability is attributed to donor-acceptor synergy across the van der Waals gap, where electron-donating $\mathrm{V}_{\mathrm{S}}$ levels and deep-acceptor $\mathrm{V}_{\mathrm{Zn}}$ levels undergo electronic coupling. Despite having a lower binding energy than the zinc pair, the $(\mathrm{V}_{\mathrm{S}} - \mathrm{V}_{\mathrm{O,int}})$ pair exhibits significantly lower formation energies (4.81~eV under anion-poor conditions). This indicates that oxygen-sulfur vacancy pairs are the thermodynamically accessible neutral species at the interface, driven primarily by low combined formation energy rather than strong cross-interface coupling.

To understand the spatial effect of defect interaction, we compared a proximity configuration of the (V$_S$ -- V$_{\text{Zn,int}}$) pair (2.67~\AA, directly below V$_S$) against a distanced interfacial model (5.61~\AA). The proximity configuration is thermodynamically less stable, showing an increase in formation energy and a corresponding reduction in binding energy of 0.1~eV relative to the distanced state. This relative instability indicates that while the WS$_2$/ZnO interface generally drives defect aggregation, immediate proximity incurs a localized energy penalty due to short-range steric repulsion. Despite this thermodynamic penalty, the electronic band structures for both configurations are nearly identical (Fig.~S9), confirming that the electronic signature of the V$_S$ -- V$_{\text{Zn}}$ interaction, specifically the energetic placement of deep-level in-gap states, remains robust and insensitive to exact atomic separation.

While neutral baseline calculations serve as a reference, defects near the CBM ($n$-type conditions) carry charges. Evaluating these charged configurations reveals how Coulomb interactions alter net pairing stability. According to charge transition levels (CTLs) in Fig.~\ref{fig:defect_level}, $\mathrm{V}_{\mathrm{Zn}}$ remains ionized throughout the fundamental bandgap and does not stabilize as neutral $\mathrm{V}_{\mathrm{Zn}}$. Near the CBM ($E_{\text{F}} = 2.3$~eV), $\mathrm{V}_{\mathrm{S}}$ acts as a doubly negative charged ($\mathrm{V}_{\mathrm{S}}''$), $\mathrm{V}_{\mathrm{Zn}}$ exists as an ionized acceptor ($\mathrm{V}_{\mathrm{Zn}}''$), and $\mathrm{V}_{\mathrm{O}}$ remains neutral ($\mathrm{V}_{\mathrm{O}}$). Thus, the $(\mathrm{V}_{\mathrm{S}} - \mathrm{V}_{\mathrm{Zn}})''''$ and $(\mathrm{V}_{\mathrm{S}} - \mathrm{V}_{\mathrm{O}})''$ pairs represent the thermodynamic states specifically in this near-CBM energy.

Evaluating these charged pairs against the neutral baseline highlights the role of electrostatics near the CBM. For $(\mathrm{V}_{\mathrm{S}} - \mathrm{V}_{\mathrm{O}})''$, pairing $\mathrm{V}_{\mathrm{S}}''$ with neutral $\mathrm{V}_{\mathrm{O}}$ introduces no net inter-site Coulombic force ($q_1 \cdot q_2 = 0$). As a result, the binding energy remains nearly identical to the neutral reference, yielding $-0.55$~eV for the interface and $-0.61$~eV for the bulk.

In contrast, for $(\mathrm{V}_{\mathrm{S}} - \mathrm{V}_{\mathrm{Zn}})''''$, pairing $\mathrm{V}_{\mathrm{S}}''$ with $\mathrm{V}_{\mathrm{Zn}}''$ introduces inter-layer Coulomb repulsion ($q_1 \cdot q_2 = +4$). This significantly weakens the binding energy to $-0.22$~eV at the interface and $-0.36$~eV in the bulk relative to the neutral reference ($-1.03$~eV and $-0.64$~eV). Notably, while the neutral baseline prefers interfacial pairing, the $(\mathrm{V}_{\mathrm{S}} - \mathrm{V}_{\mathrm{Zn}})''''$ state near the CBM exhibits stronger binding in bulk $\mathrm{ZnO}$. This trend reversal occurs because larger separation between $\mathrm{V}_{\mathrm{S}}$ and $\mathrm{V}_{\mathrm{Zn,bulk}}$, combined with higher dielectric screening in bulk $\mathrm{ZnO}$, mitigates the repulsive electrostatic relative to the interfacial pair.

Importantly, the formation energy of these charged species near the CBM are significantly lower. For $(\mathrm{V}_{\mathrm{S}} - \mathrm{V}_{\mathrm{Zn,int}})''''$, $E_{\text{f}}$ drops to 2.61~eV under anion-poor conditions (5.75~eV under anion-rich conditions). As a result, $(\mathrm{V}_{\mathrm{S}} - \mathrm{V}_{\mathrm{Zn}})''''$ becomes the most thermodynamically favorable defect pair at the $n$-type interface, forming more readily than both the neutral $(\mathrm{V}_{\mathrm{S}} - \mathrm{V}_{\mathrm{O,int}})$ (4.81~eV) and the $(\mathrm{V}_{\mathrm{S}} - \mathrm{V}_{\mathrm{O,int}})''$ pair (6.34~eV). Thus, even though inter-layer Coulomb repulsion reduces its binding energy, the low formation energy establishes that the $-4$ pair remains the dominant defect cluster at the heterojunction.

In short, near the CBM, the $(\mathrm{V}_{\mathrm{S}} - \mathrm{V}_{\mathrm{Zn}})''''$ cluster dominates the interface due to its low formation energy (2.61~eV), despite Coulomb repulsion weakening its binding energy relative to the neutral state. By contrast, $\mathrm{V}_{\mathrm{S}} - \mathrm{V}_{\mathrm{O}}$ pairs show stable binding across all charge states. Because $(\mathrm{V}_{\mathrm{S}} - \mathrm{V}_{\mathrm{Zn}})$ introduces deep in-gap states, its high concentration directly limits the internal quantum efficiency of the device. This makes targeted chemical passivation essential to remove these primary recombination centers.

By comparing the single-defect band structures with combined configurations, we also observe that the interfacial environment induces specific energy shifts and orbital hybridization that are absent in remote positions (Fig.~S15). 

A primary result of this analysis is the change in the oxygen vacancy trap state when coupled with a sulfur vacancy at the interface ($V_S - V_{O\text{-inter}}$). While the isolated interfacial oxygen vacancy introduces an occupied mid-gap level at $\approx 0.75$~eV, the presence of the neighboring V$_S$ induces an upward energy shift of approximately $0.05$~eV, pushing the state further into the gap. Furthermore, the deep unoccupied states of V$_S$, originally situated at $1.8$~eV and $2.2$~eV, exhibit increased dispersion and splitting in this combined model. This behavior suggests a degree of hybridization between metal orbitals pointing into the the $O$ and $S$ vacancy sites.

The interaction between the interfacial sulfur vacancy and the zinc vacancy ($V_S - V_{Zn\text{-inter}}$) reveals a different trend. The isolated $V_{Zn\text{-inter}}$ state, a shallow unoccupied acceptor initially located $\approx 0.15$~eV above the VBM, undergoes a downward energy shift of approximately $0.04$~eV in the presence of V$_S$. This shift pulls the V$_{Zn}$ level closer to the valence band edge, narrowing the activation energy required for hole generation. This suggests that $S$-deficiency at the interface might actually enhance the p-type potential of the heterostructure by facilitating easier electronic excitation into the $Zn$ vacancy sites.

In contrast to the shifts observed at the junction, defect states located within bulk ZnO layer appear to be unperturbed. In the $V_S - V_{O\text{-bulk}}$ and $V_S - V_{Zn\text{-bulk}}$ configurations, the shallow donor and acceptor levels remain relatively the same, with negligible shifts from their individual positions at $0.05$~eV and $0.08$~eV, respectively. The spatial separation between these bulk sites and the interfacial sulfur vacancy prevents significant electrostatic interaction or orbital overlap.

\begin{figure}[hb!]
    \centering
    \includegraphics[height=0.85\textheight, width=\textwidth, keepaspectratio]{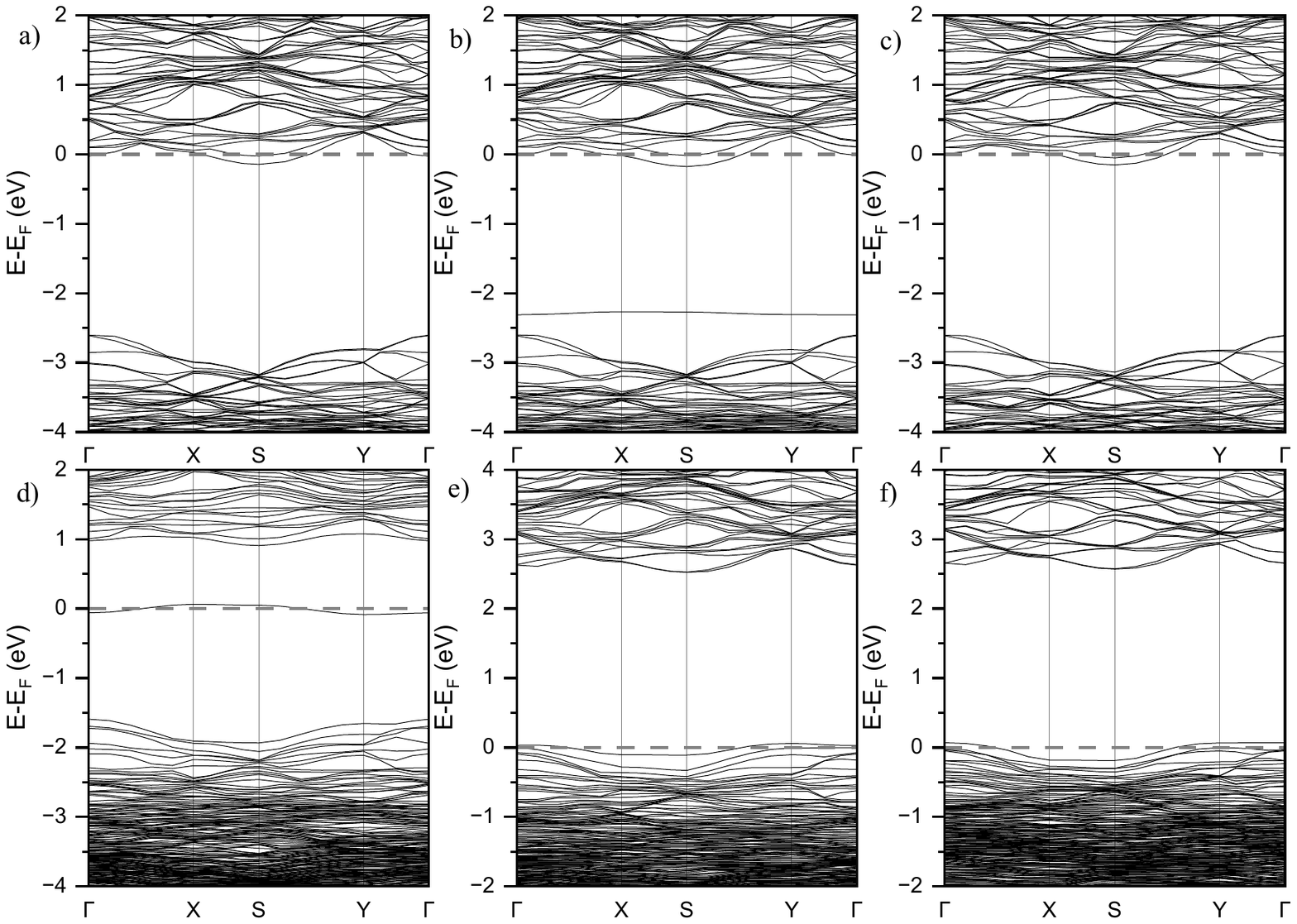}
    \caption{Calculated electronic band structures for interstitial hydrogen (H$_i$) and its corresponding defect complexes. The figure shows a) isolated H$_i$, b) H$_i$ coupled with an interfacial oxygen vacancy ($H_i + V_{\text{O,inter}}$), c) bulk oxygen vacancy ($H_i + V_{\text{O,bulk}}$), d) sulfur vacancy ($H_i + V_{\text{S}}$), e) interfacial zinc vacancy ($H_i + V_{\text{Zn,inter}}$), and f) bulk zinc vacancy ($H_i + V_{\text{Zn,bulk}}$). In figures a)–c), the donor-dominated interactions push the Fermi level near the conduction band minimum. In contrast, figures e) and f) the donated electron directly populates pre-existing shallow acceptor states. The Fermi level is set to zero in all plots.}
    \label{fig:hydrogen_interstitial}
\end{figure}

Across all combined models, it is important to note that there is a preservation of defect occupancy and chemical character. Despite the energy shifts, V$_O$ states remain consistently occupied (donors), while V$_{Zn}$ and V$_S$ states remain unoccupied (acceptors). This indicates that the nature of these vacancies, whether they contribute carriers or trap them, is an intrinsic property. 

\subsubsection{Interstitial Hydrogen and Vacancy Pairs}

As mentioned briefly in the earlier section, n-type conductivity is often observed for ZnO\cite{janotti_native_2007}. While there is growing evidence that ionized defects in the substrate are the source of the electrons populating the conduction band, we are in position to specify this mechanism by explicit calculations of WS$_2$/ZnO. The bond-centered hydrogen interstitial ($H_{i\text{-BC}}$) induces a shift of the Fermi level toward the conduction band in our DFT calculations (Fig.~\ref{fig:hydrogen_interstitial}a), consistent with the idea that $H_{i\text{-BC}}$ is a major source of n-type doping in ZnO. This confirms that unintentional hydrogen incorporation can serve as the dominant donor species also in the heterostructures. For their application in LEDs, this provides the required carrier density for electron injection from the ZnO into the WS$_2$ active layer. Interstitial hydrogen (H$_i$) interacts with native structural vacancies to control the charge balance and Fermi level of the heterostructure. How these defect complexes behave depends entirely on whether the original vacancy acts as a donor or an acceptor.

When H$_i$ pairs with native donor vacancies like V$_S$ or V$_O$, the result is a dynamic dual-donor complex. For instance, pairing H$_i$ with a sulfur vacancy completely alters the local electronic structure (Fig.~\ref{fig:hydrogen_interstitial}d). An isolated V$_S$ creates two unoccupied states at 1.9~eV and 2.2~eV above the valence band maximum (VBM). However, after introducing H$_i$ to the ZnO, a single partially occupied state remains around 1.5~eV above the VBM. Oxygen vacancies behave similarly (Fig.~\ref{fig:hydrogen_interstitial}b-c): pairing V$_O$ with H$_i$ introduces excess electrons that push the Fermi level up to the conduction band minimum (CBM), making the system strongly $n$-type.

Contrasting behavior is found by combining H$_i$ with $V_{\text{Zn}}$ because $V_{\text{Zn}}$ is a native acceptor (Fig.~\ref{fig:hydrogen_interstitial}e-f). An isolated $V_{\text{Zn}}$ creates shallow states just above the VBM (see Fig.~\ref{fig:defect_level}a). When H$_i$ is introduced, the electron it donates does not create a new mid-gap trap or push the Fermi level across the band gap. Instead, it drops directly into the existing shallow acceptor states left by the zinc vacancy. 

This electronic compensation mechanism is rather evident from the data: the Fermi level barely shifts ($+0.04$~eV in the bulk and $+0.07$~eV at the interface), while the occupancy of the band edges increases significantly. At the interface, for example, the main acceptor band goes from 0.22 to 0.51 occupancy. The complex absorbs the extra electron while keeping the energy state pinned to the VBM.

Ultimately, the $H_i + V_{\text{Zn}}$ pair functions as a polarized shallow acceptor complex. They change the local charge environment through compensation without disrupting the underlying band structure of the host lattice.

\section{Conclusion}

Our facet screening identifies the non-polar ($10\overline{1}0$) $m$-plane as the preferred substrate surface to maintain direct Type-I band alignment in WS$_2$/ZnO heterostructures. Beyond structural interface matching, junction performance is strongly governed by the underlying defect thermodynamics. While this non-polar surface sets the ideal baseline alignment, operational performance is also determined by the specific native defects that form during growth.

For isolated defects under anion-poor growth conditions, sulfur vacancies and interfacial oxygen vacancies exhibit the lowest formation energies, making them the most prevalent point defects at the interface and introducing deep-level trap states into the bandgap. Meanwhile, isolated zinc vacancies act as shallow acceptors due to the Type-I band alignment, providing acceptor state near the valence band edge. As the Fermi level shifts toward the conduction band under $n$-type conditions, these isolated defects change charge states and drive defect pair formation across the gap.

Charge transition level analysis indicates that near the conduction band minimum ($n$-type regime), $\mathrm{V}_{\mathrm{S}}''$ and $\mathrm{V}_{\mathrm{Zn}}''$ stabilize in their fully ionized $-2$ states ($\mathrm{V}_{\mathrm{S}}''$ and $\mathrm{V}_{\mathrm{Zn}}''$), whereas V$_O$ remains neutral (V$_O$). Pairing these charged defects alters interfacial stability across the van der Waals gap. Because V$_O$ remains neutral, (V$_S$ - V$_O$)'' pairs maintain negative binding energies comparable to the neutral state. In contrast, inter-layer Coulomb repulsion between $\mathrm{V}_{\mathrm{S}}''$ and $\mathrm{V}_{\mathrm{Zn}}''$ weakens the binding energy of $(\mathrm{V}_{\mathrm{S}} - \mathrm{V}_{\mathrm{Zn}})''''$. Nevertheless, at $E_{\text{F}} = 2.3$~eV, the strong $qE_{\text{F}}$ contribution lowers the formation energy of $(\mathrm{V}_{\mathrm{S}} - \mathrm{V}_{\mathrm{Zn}})''''$ to 2.61~eV under anion-poor conditions. This low formation energy makes the $-4$ cluster the most thermodynamically abundant defect pair at the $n$-type interface, despite its weaker binding energy.

In addition to intrinsic native defect pairs, extrinsic hydrogen impurities introduced during processing further modify the interfacial electronic structure. Hydrogen interstitials (H$_i$) act as shallow $n$-type donors, but can readily interact with native zinc vacancies (V$_{Zn}$). Rather than introducing deep trap states, the donor electron compensates the shallow acceptor level of V$_{Zn}$, preserving the underlying band-edge alignment of the heterojunction.

Translating these thermodynamic insights into device performance requires suppressing these primary non-radiative pathways during film growth. Maximizing the efficiency of 2D/3D hybrid LEDs requires controlling anion vacancy formation during synthesis. Suppressing isolated vacancies and applying targeted chemical passivation strategies to eliminate the dominant $(\mathrm{V}_{\mathrm{S}} - \mathrm{V}_{\mathrm{Zn}})''''$ clusters remain necessary steps to remove non-radiative recombination centers at the active interface.

\section{Methods}

First-principles calculations were performed using spin-polarized density functional theory (DFT) as implemented in the VASP package (version 6.3.0) \cite{Kresse.1996, Kresse.1999}. Ion--valence electron interactions were described using the projector augmented-wave (PAW) method \cite{Blochl.1994}, with a plane-wave cutoff energy of 500~eV. Van der Waals dispersion forces were accounted for using Grimme's DFT-D3 scheme with Becke--Johnson damping \cite{Grimme.2011}. Orbital-projected band structures were extracted using the VASPKIT code \cite{Wang.2021b}, and defect formation energies as well as charge transition levels were evaluated using the \texttt{doped} Python package \cite{Kavanagh2024}.

\subsection*{Interface Facet Screening}
For interface facet screening, heterostructure interface models as shown in Fig.~\ref{fig:m_plane}, \ref{fig:a_plane} and \ref{fig:polar} were constructed by placing a $\text{WS}_2$ monolayer on a 6-layer $\text{ZnO}$ substrate. Unphysical polar surface states were eliminated by passivating the bottom layer of the $\text{ZnO}$ substrate with pseudo-hydrogen atoms. A 20~\AA\ vacuum layer was applied normal to the interface to prevent spurious interactions between periodic images. Brillouin zone sampling was carried out using an $8 \times 5 \times 1$ Monkhorst-Pack $k$-point grid. Exchange-correlation effects were treated with a modified HSE06 hybrid functional \cite{heyd2003hybrid, krukau2006influence} using a 37.5\% fraction of exact exchange ($\alpha = 0.375$). Spin-orbit coupling (SOC) was explicitly included for all facet screening calculations.

\subsection*{Interfacial Defect Calculations}
For defect thermodynamics, a sandwiched $\text{WS}_2/\text{ZnO}$ heterostructure was employed without a vacuum region. Defect modeling was performed in a $3 \times 3 \times 1$ supercell ($\approx 270$ atoms) to minimize artificial inter-defect interactions across periodic boundaries, sampled with a $2 \times 2 \times 1$ Monkhorst-Pack $k$-point grid. To manage the computational cost of the 270-atom hybrid functional calculations, SOC was omitted while maintaining the same exchange-correlation and DFT-D3 setup.

\begin{acknowledgement}

The authors gratefully acknowledge financial support from the DFG within the IRTG 2803: 2D Mature, project No. 461605777. Additionally, the authors acknowledge the computing time granted by the Center for Computational Sciences and Simulation (CCSS) of the Universität of Duisburg-Essen and provided on the supercomputer amplitUDE (DFG project 459398823; grant ID INST 20876/423-1 FUGG) at the Zentrum für Informations- und Mediendienste (ZIM).

\end{acknowledgement}

\begin{suppinfo}

The Supporting Information is available at DOI/xxx.

\end{suppinfo}

\bibliography{main}
\clearpage


\section*{Supplementary material}

# Supporting Information

## Facet-Dependent Electronic Properties and Interfacial Point Defect Interactions in $WS_2$/ZnO Heterostructures

Dedi Sutarma, Peter Kratzer

Department of Physics, University of Duisburg-Essen, Lotharstr. 1, Duisburg, 47048, Germany

E-mail: dedi.sutarma@uni-due.de; peter.kratzer@uni-due.de

**Band Structure of Pristine $WS_2$**

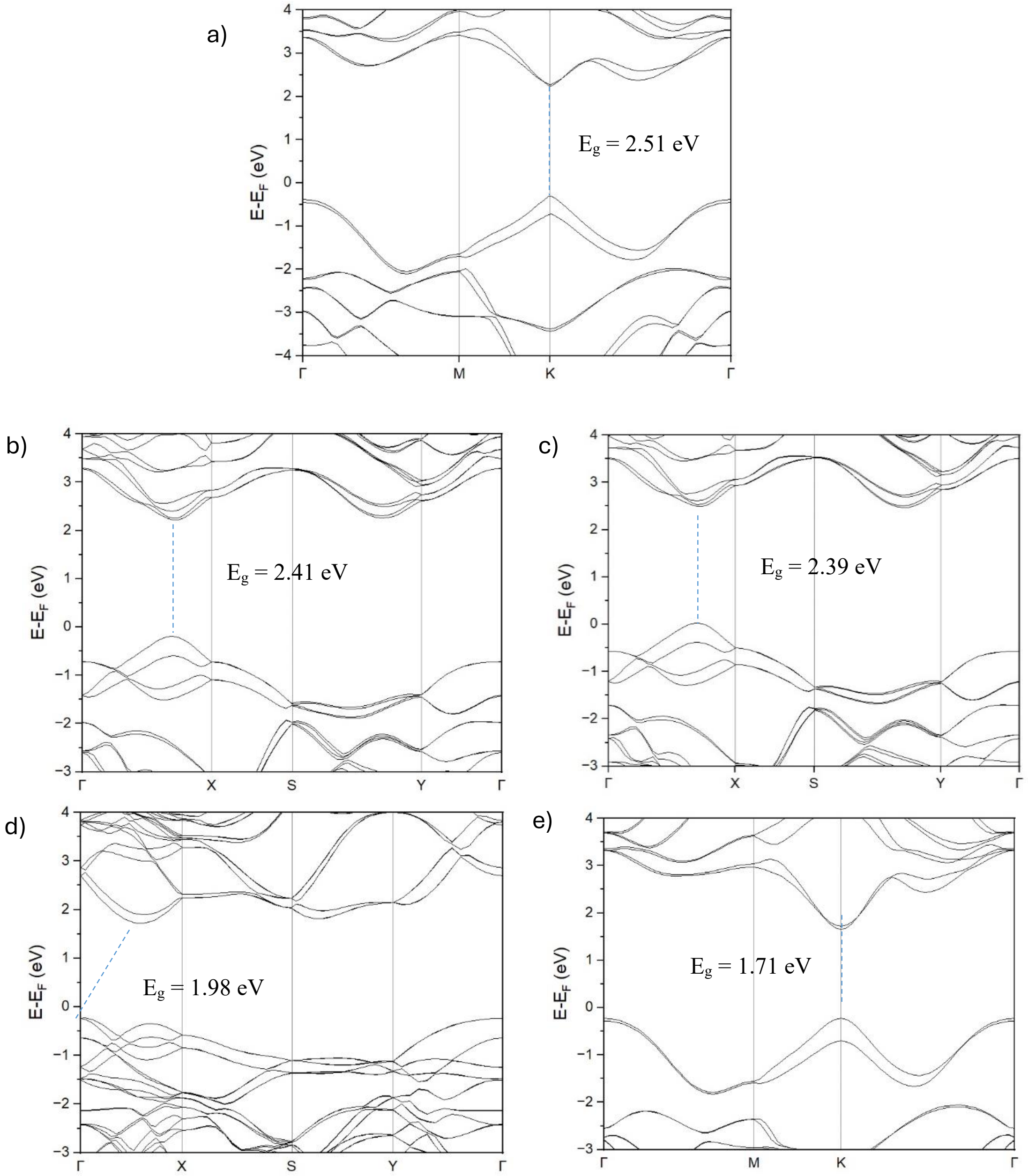


Figure S1. Evolution of the $WS_2$ electronic band structure under substrate-induced strain. Comparison of isolated a) $WS_2$ and strained $WS_2$ matched to various ZnO facets: b) $1\bar{1}00$, c) $1\bar{1}00$ double-sided, d) $1\bar{2}00$, and e) 0001. The bands are calculated with HSE ($\mu = 0.2$, $\alpha = 0.375$)

**ZnO Substrate**

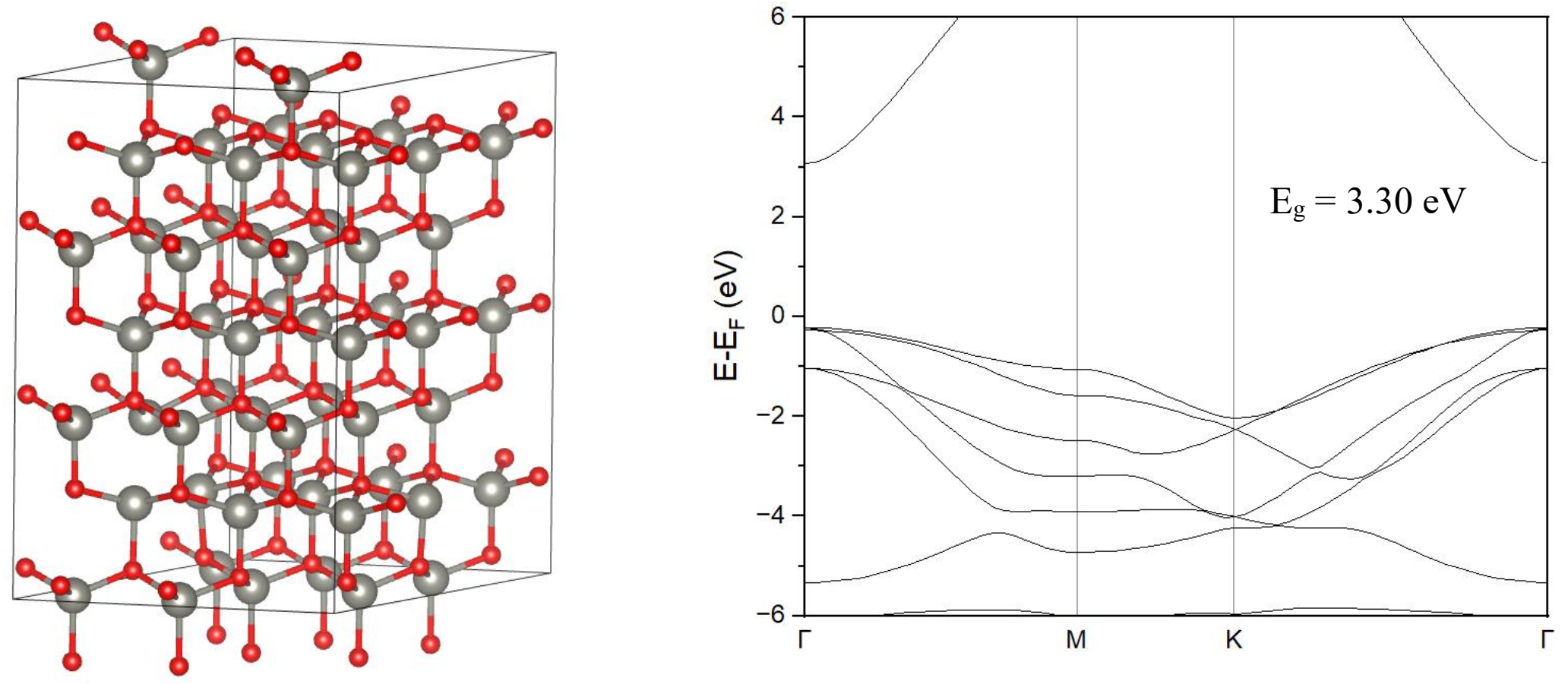


Figure S2. ZnO bulk structure and calculated band structure. The band is calculated with HSE (μ = 0.2, α = 0.375).

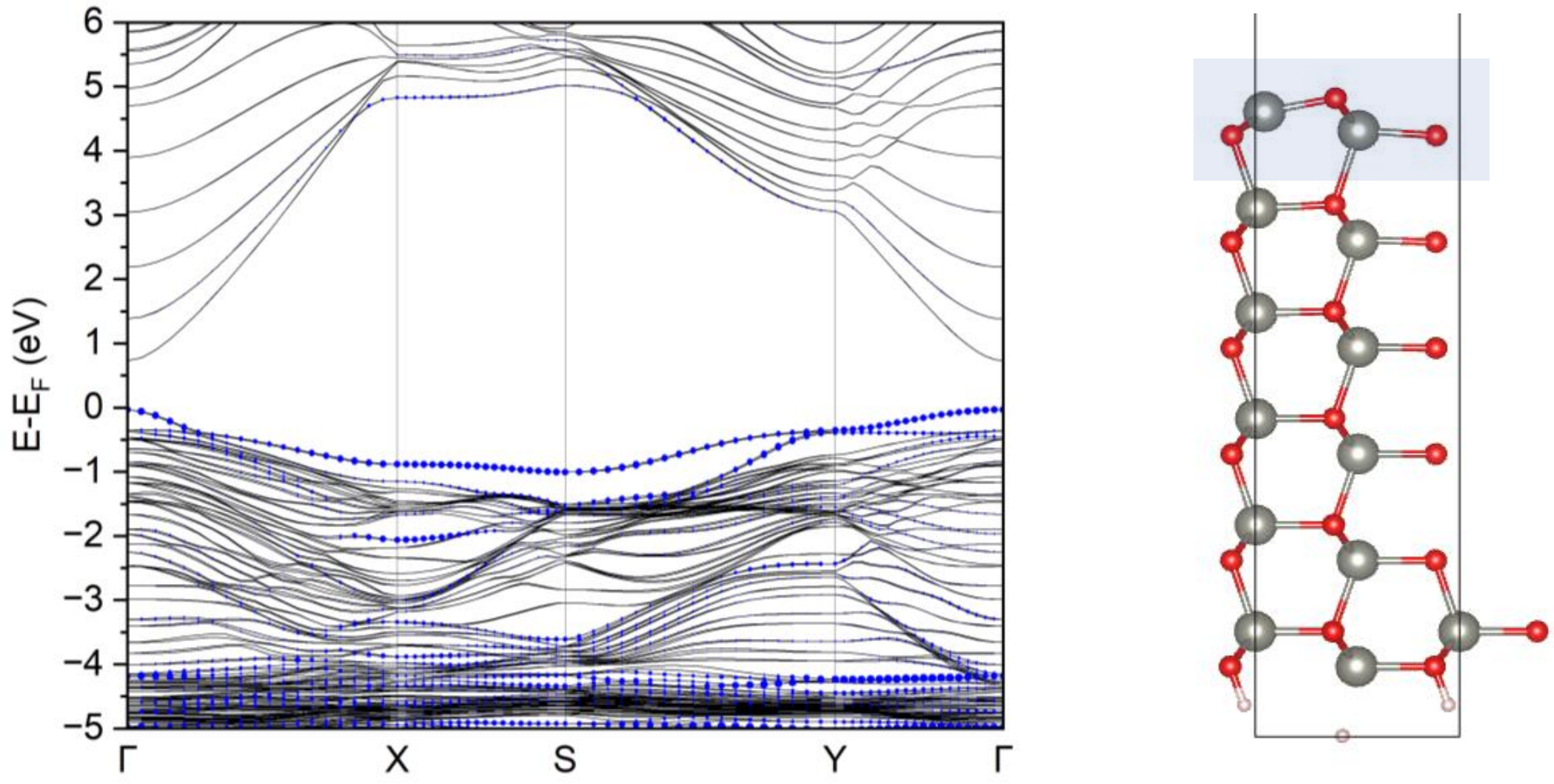


Figure S3. Height-resolved band structure reveals surface relaxation induced band bending on ZnO $1\bar{1}00$. The bulk Zn–O–Zn bond angle of 108.0° compresses significantly to 60.7° at the relaxed surface, driving the electronic shifts. The electronic structure was evaluated at the PBE level of theory.

**$1\bar{1}00$ double-sided**

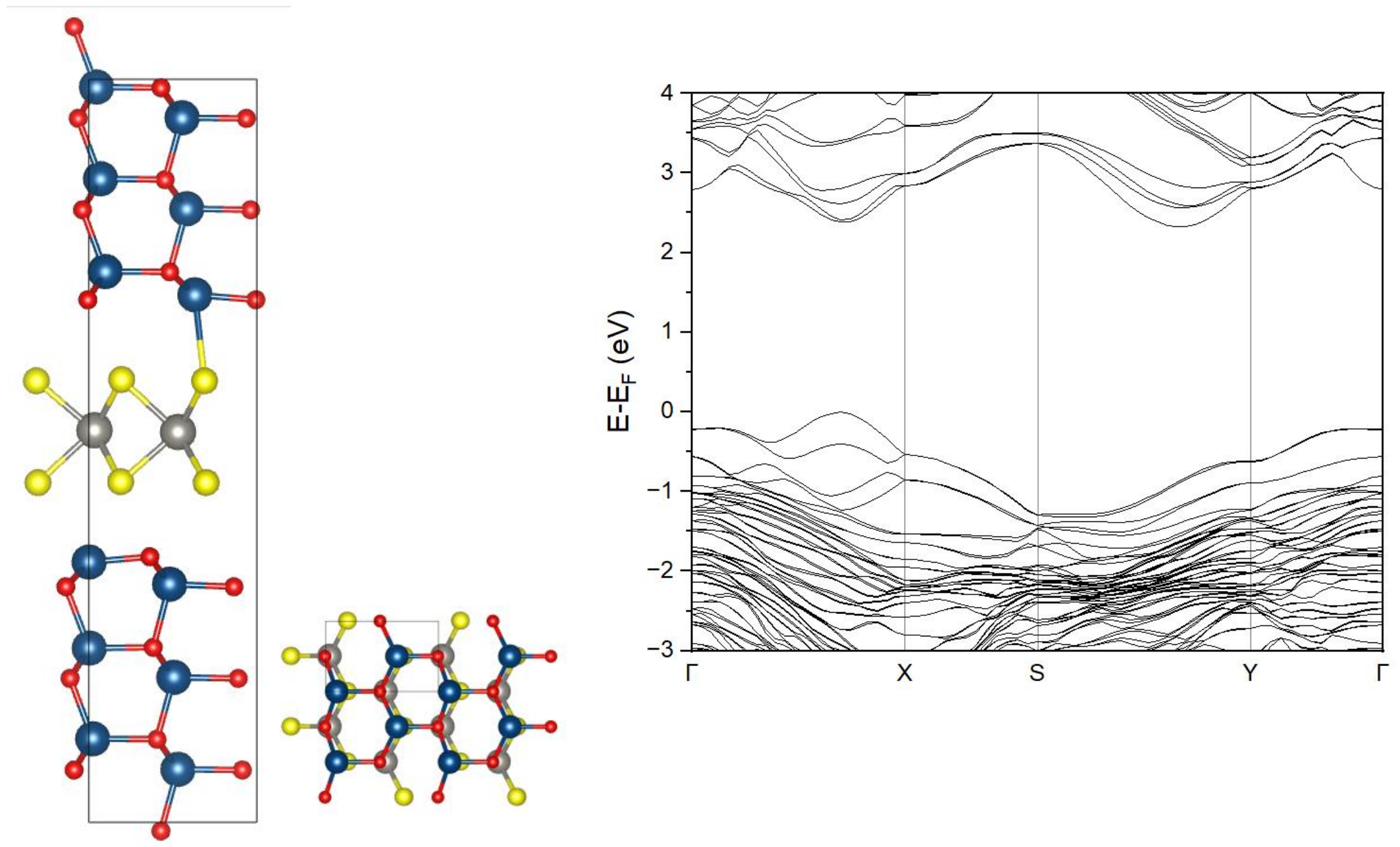


Figure S4. Double-sided ZnO $1\bar{1}00$ model and band structure, matching the single-sided configuration. The band was calculated with HSE ($\mu = 0.2$, $\alpha = 0.375$).

**Defect Formation Energy**

The thermodynamic stability and equilibrium concentrations of point defects, hydrogen complexes, and interfacial substitutions at the $WS_2$/ZnO heterostructure are evaluated using first-principles total energy calculations. The formation energy ($E_f$) of a given defect or interfacial complex in charge state q is modeled as a function of the electron chemical potential (Fermi level, $\varepsilon_F$) and the relative chemical potentials of the constituent atomic species ($\Delta\mu_i$) according to the expression:

$$E_f(D^q) = E_{tot}(D^q) - E_{tot}(\text{pristine}) - \sum n_i \mu_i + q(\varepsilon_F + \varepsilon_V) + \Delta_{corr}$$

where $E_{tot}(D^q)$ represents the total energy of the supercell containing the defect configuration in charge state q, and $E_{tot}$ (pristine) is the total energy of the corresponding fully relaxed pristine heterostructure supercell. The integer n_i denotes the number of atoms of species i (i = W, S, Zn, O, H) added ($n_i > 0$) or removed ($n_i < 0$) to construct the defect site. The Fermi level $\varepsilon_F$ is referenced to the valence band maximum (VBM) of the pristine system, $\varepsilon_V$, and $\Delta_{corr}$ represents the FNV (Freysoldt, Neugebauer, and Van de Walle) correction scheme implemented to mitigate artificial periodic electrostatic interactions between charged periodic images. The process was carried out in doped code.

The atomic chemical potentials ($\mu_i$) vary depending on the experimental growth and environmental conditions. To simplify the analysis, we define these potentials relative to the total energy per atom of their respective elemental ground states ($\mu_{i,bulk}$):

$\mu_i = \mu_{i,bulk} + \Delta\mu_i$

Under equilibrium conditions, the relative chemical potentials ($\Delta\mu_i$) must be less than or equal to zero ($\Delta\mu_i \leq 0$) to prevent the spontaneous precipitation of pure elemental phases. Using identical hybrid functional parameters ($\mu$ = 0.2 bohr $^{-1}$, $\alpha$ = 0.375), the reference energies per atom were calculated as follows:

Table S1. Chemical potential of elements in heterostructure

| Element | μ (eV) | Reference Phase |
|---|---|---|
| Tungsten ($\mu_W$) | 16.37 | body-centered cubic bulk W |
| Zinc ($\mu_{Zn}$) | 2.79 | hexagonal close-packed bulk Zn |
| Sulfur ($\mu_S$) | 5.81 | orthorhombic $S_8$ |
| Oxygen ($\mu_O$) | 8.02 | isolated $O_2$ molecule |
| Hydrogen ($\mu_H$) | 5.69 | isolated $H_2$ molecule |

Under equilibrium conditions, the coexistence of the $WS_2$ monolayer and the ZnO substrate requires that the chemical potentials are simultaneously constrained by the calculated formation enthalpies ($\Delta H_f$) of the parent bulk compounds.

$\Delta\mu_W + 2\Delta\mu_S = \Delta H_f(WS_2) = -2.029$ eV

$\Delta\mu_{Zn} + \Delta\mu_O = \Delta H_f(ZnO) = -2.125$ eV

To capture the thermodynamics of the system, two distinct physical growth limits are established based on the host constraints:

(i) Anion-Poor (Metal-Rich):

This region represents synthesis environments heavily saturated with the constituent metals, forcing $\Delta\mu_W = 0$ eV and $\Delta\mu_{Zn} = 0$ eV. Under these boundary parameters, the corresponding lower limits for the sulfur and oxygen phases are -1.015 eV and -2.125 eV $\Delta\mu_S$ and $\Delta\mu_O$, respectively.

(ii) Anion-Rich (Metal-Poor):

This region represents maximum saturation of the non-metal species, fixing the anion potentials at their absolute elemental upper bounds: $\Delta\mu_S = 0$ eV and $\Delta\mu_O = 0$ eV. Consequently, the metal potentials drop to their lowest permissible bounds, -2.029 eV and -2.125 eV for $\Delta\mu_W$ and $\Delta\mu_{Zn}$, respectively.

To ensure the heterostructure remains stable against decomposition into unwanted binary or ternary compounds, the relative chemical potentials must be limited by the formation enthalpies of these competing phases. Based on our HSE level of theory, we give the corresponding compound formation enthalpies:

Table S2. Secondary phase formation enthalpies of $WS_2$/ZnO

| **Secondary Phase** | **$\Delta H_f$ (eV)** |
|---|---|
| ZnS | -1.950 |
| $WO_3$ | -3.520 |
| $ZnWO_4$ | -8.120 |

To project these constraints onto a two-dimensional stability plot with $\Delta\mu_O$ on the x-axis and $\Delta\mu_S$ on the y-axis, we eliminate the metal variables for each competing phase as follows:

To prevent bulk ZnS from forming, the chemical potentials must satisfy:

$\Delta\mu_{Zn} + \Delta\mu_S \leq \Delta H_f(ZnS)$

Substituting the host-locked definition of $\Delta\mu_{Zn}$:

$(-2.125\text{eV} - \Delta\mu_O) + \Delta\mu_S \leq -1.950$ eV

Grouping the variables on the left gives the final boundary limit:

$\Delta\mu_S - \Delta\mu_O \leq 0.175$ eV

Similar treatment for $WO_3$ and $ZnWO_4$ yielded:

$WO_3$ : $3\Delta\mu_O - 2\Delta\mu_S \leq -1.491$ eV

$ZnWO_4$: $3\Delta\mu_O - 2\Delta\mu_S \leq -3.966$ eV

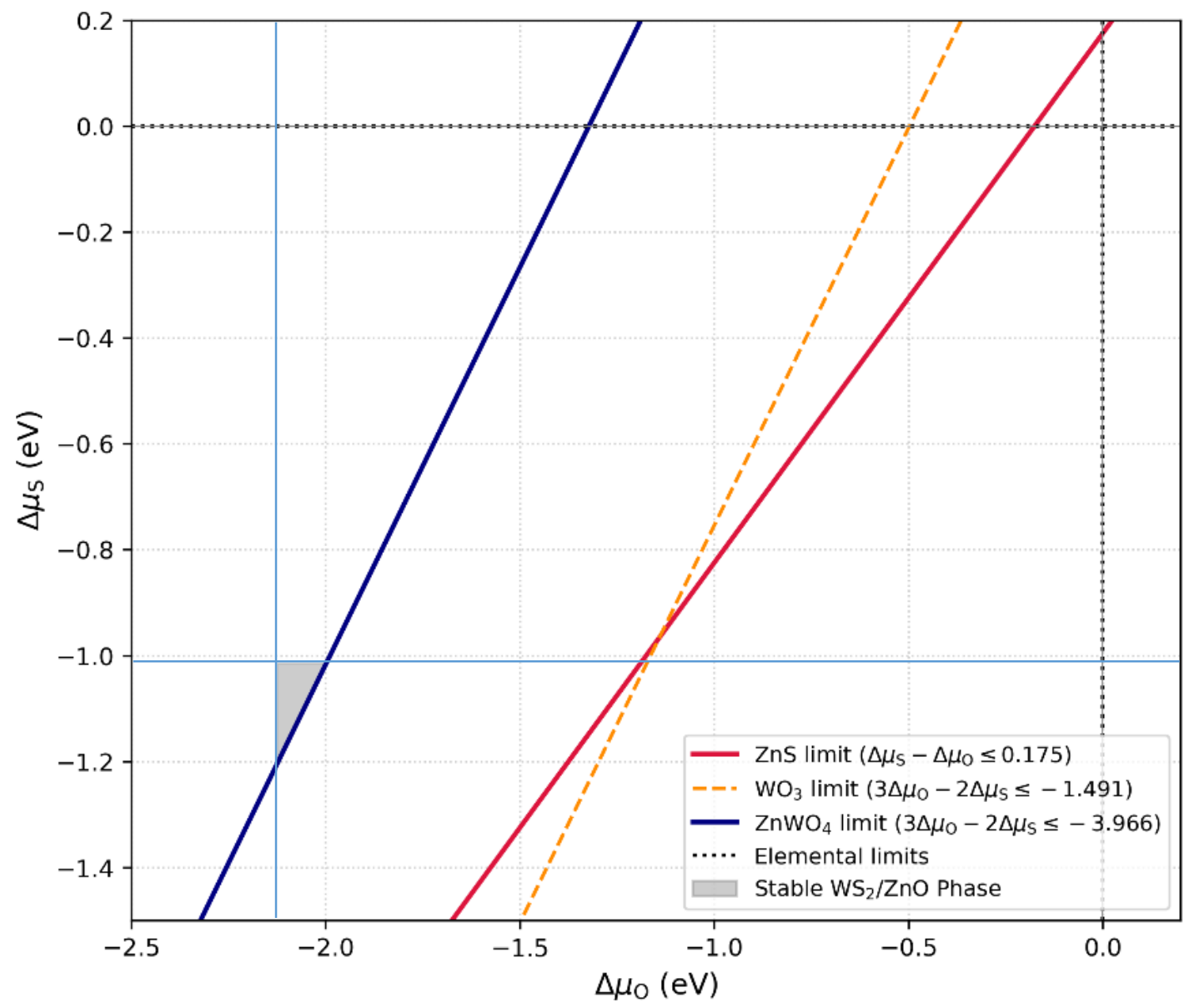


Figure S5. Thermodynamic phase diagram for the $WS_2$/ZnO heterostructure as a function of the anion chemical potentials, Δμ_O (x-axis) and Δμ (y-axis), calculated using the HSE (μ = 0.2, α = 0.375).

The calculated 2D phase stability diagram reveals that the pristine, stoichiometric $WS_2$/ZnO heterostructure is restricted to a narrow, localized chemical potential envelope deep within the anion-poor growth regime. Crucially, the thermodynamic boundary line for $ZnWO_4$ sits entirely below $WO_3$ line relative to the origin, meaning that any uncompensated increase in the oxygen chemical potential will drive the interfacial tungsten to react directly with the underlying substrate to form zinc tungstate long before pure tungsten oxides can precipitate. This encapsulation by secondary phases—$ZnWO_4$ from above and ZnS from the lateral sulfur-rich boundary—provides theoretical evidence as to why the direct, high-temperature chemical vapor deposition (CVD) of $WS_2$ monolayers onto ZnO substrate surfaces is unfavourable. Instead, by implementing a two-step fabrication strategy where the $WS_2$ monolayer is synthesized independently on a sacrificial substrate at high temperatures and subsequently transferred onto the ZnO surface near room temperature, the system successfully circumvents these severe equilibrium constraints.

## Band Structure of Point Defect

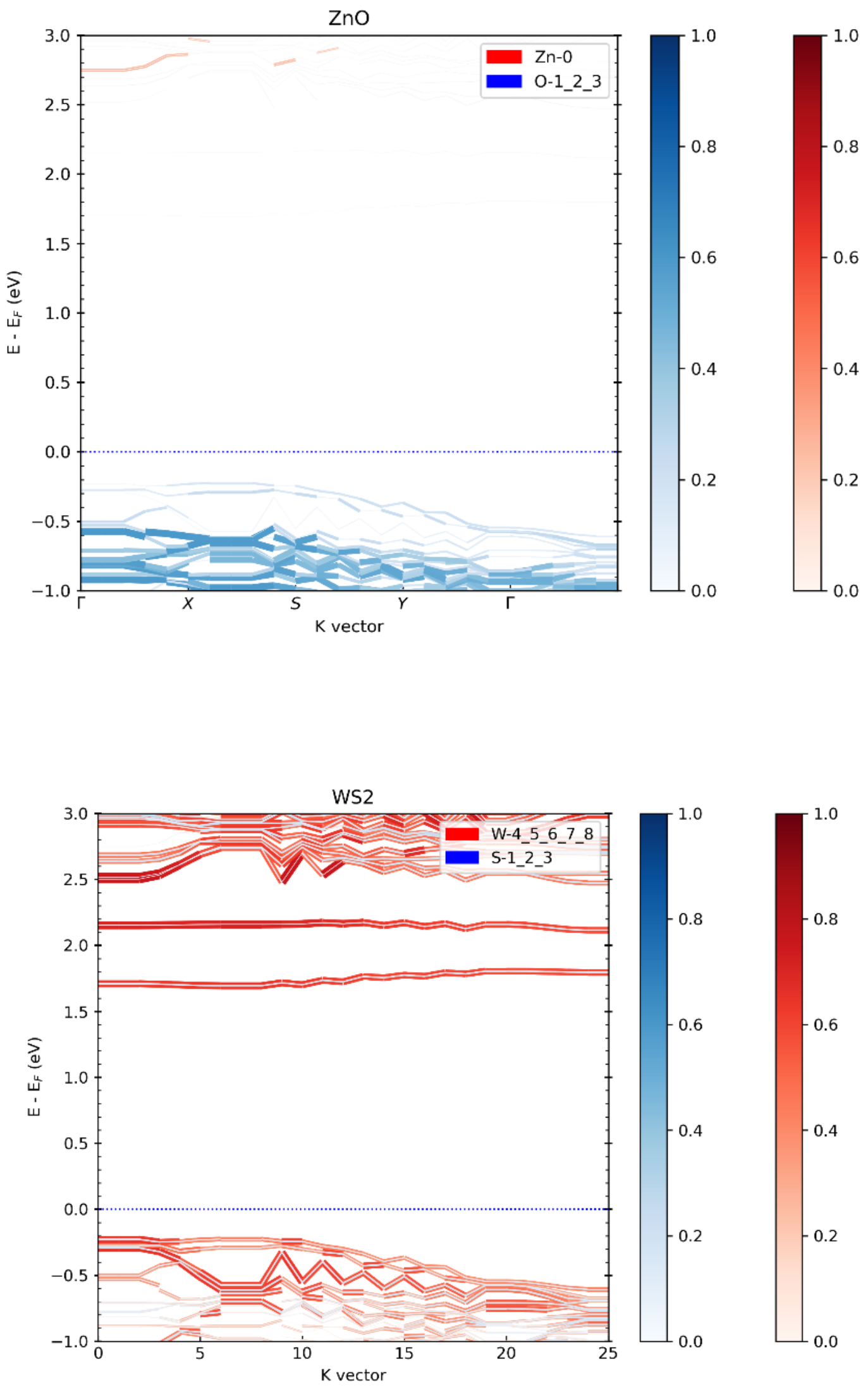


Figure S6. Projected band structure of $WS_2$/ZnO with $V_S$

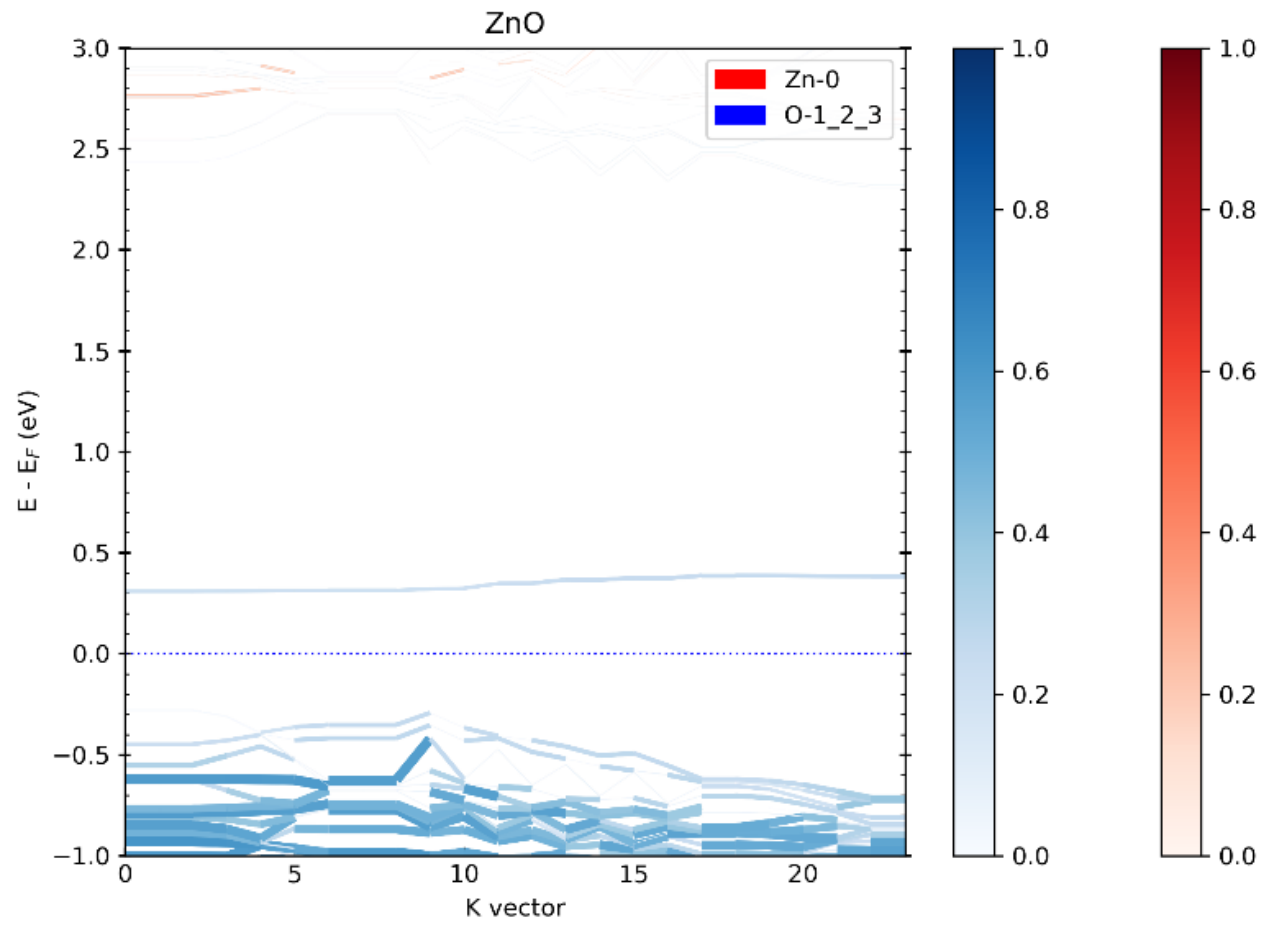

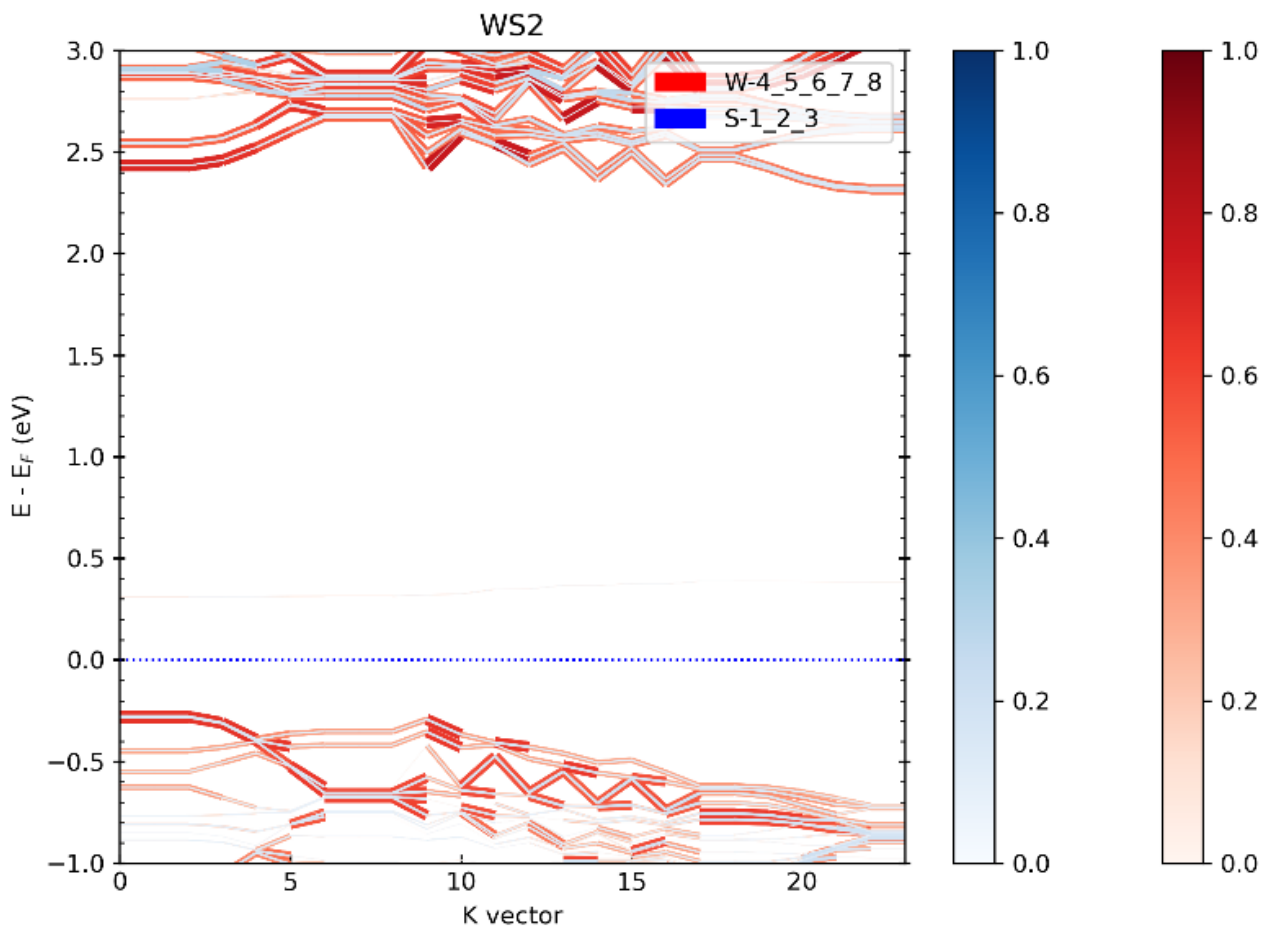


Figure S7. Projected band structure of $WS_2$/ZnO with $V_O$ inter

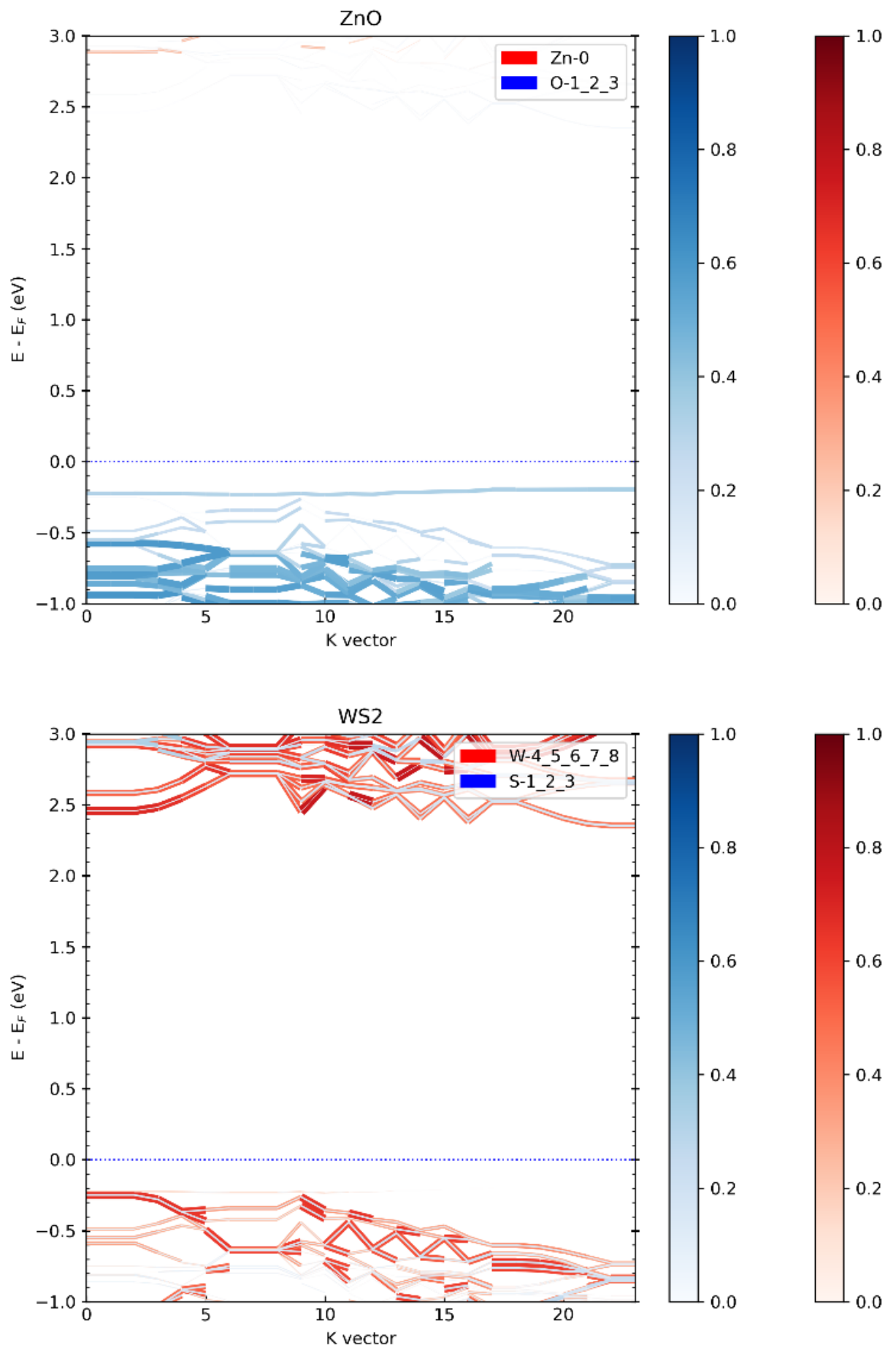


Figure S8. Projected band structure of $WS_2$/ZnO with $V_O$ bulk

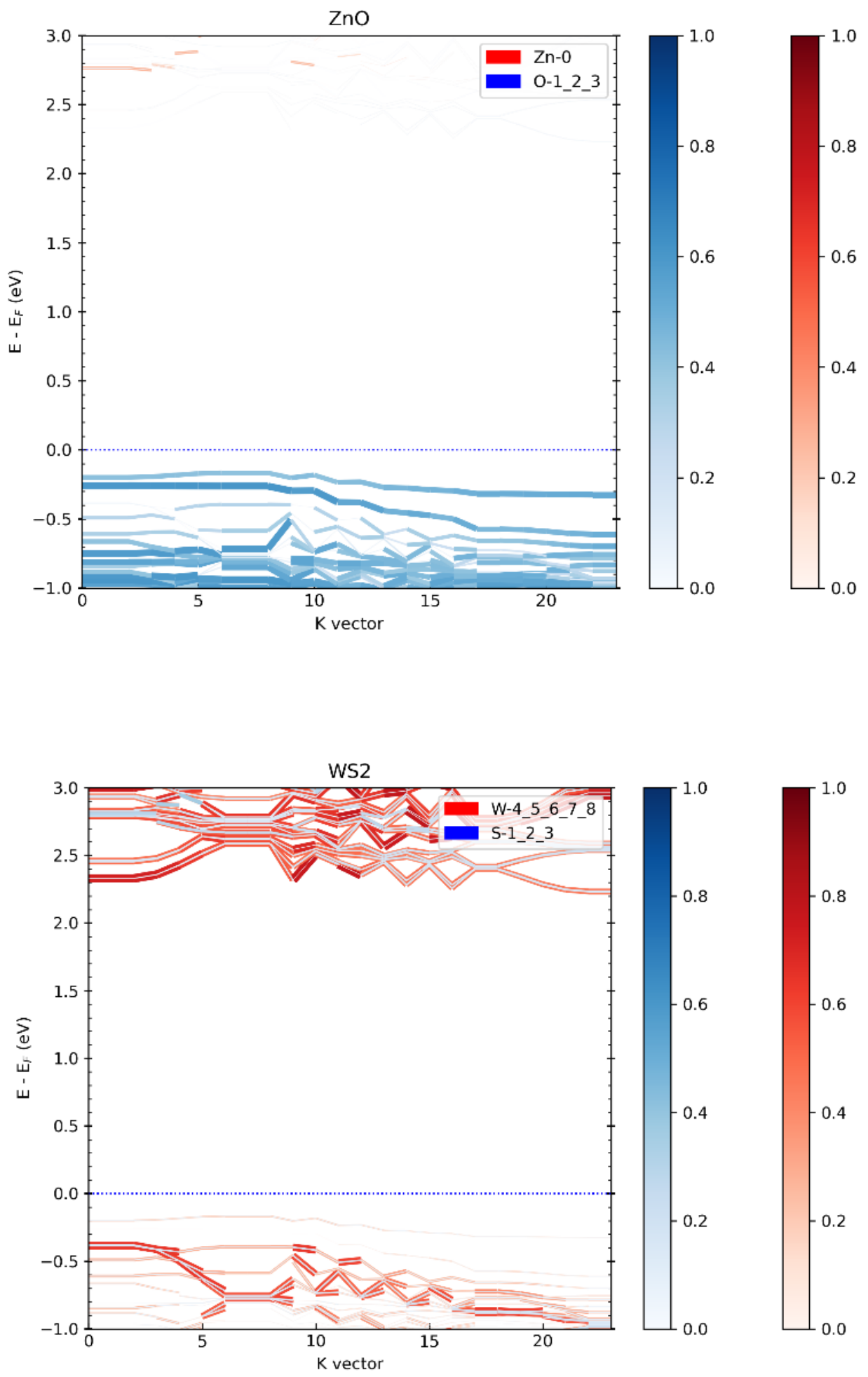


Figure S9. Projected band structure of $WS_2$/ZnO with $V_{Zn}$ inter

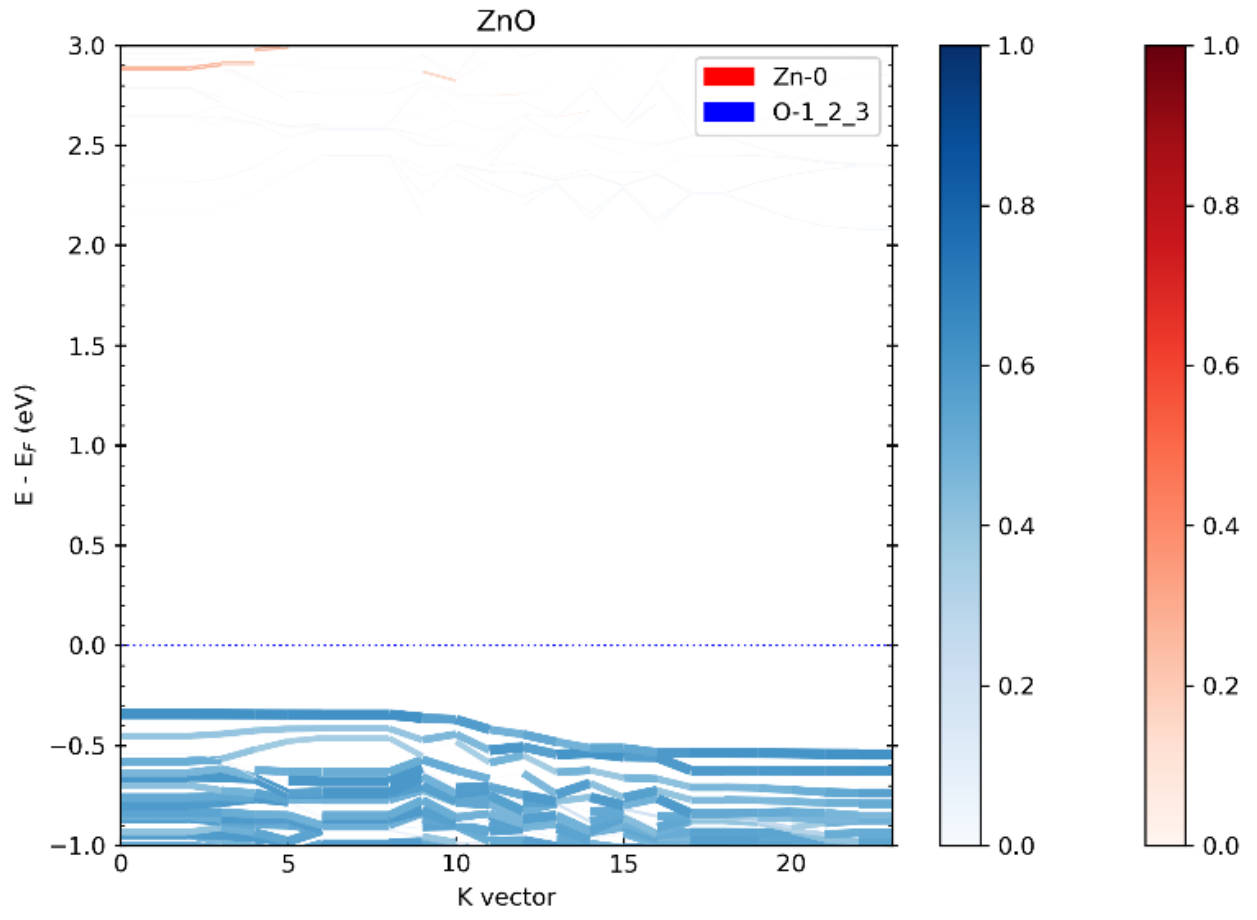

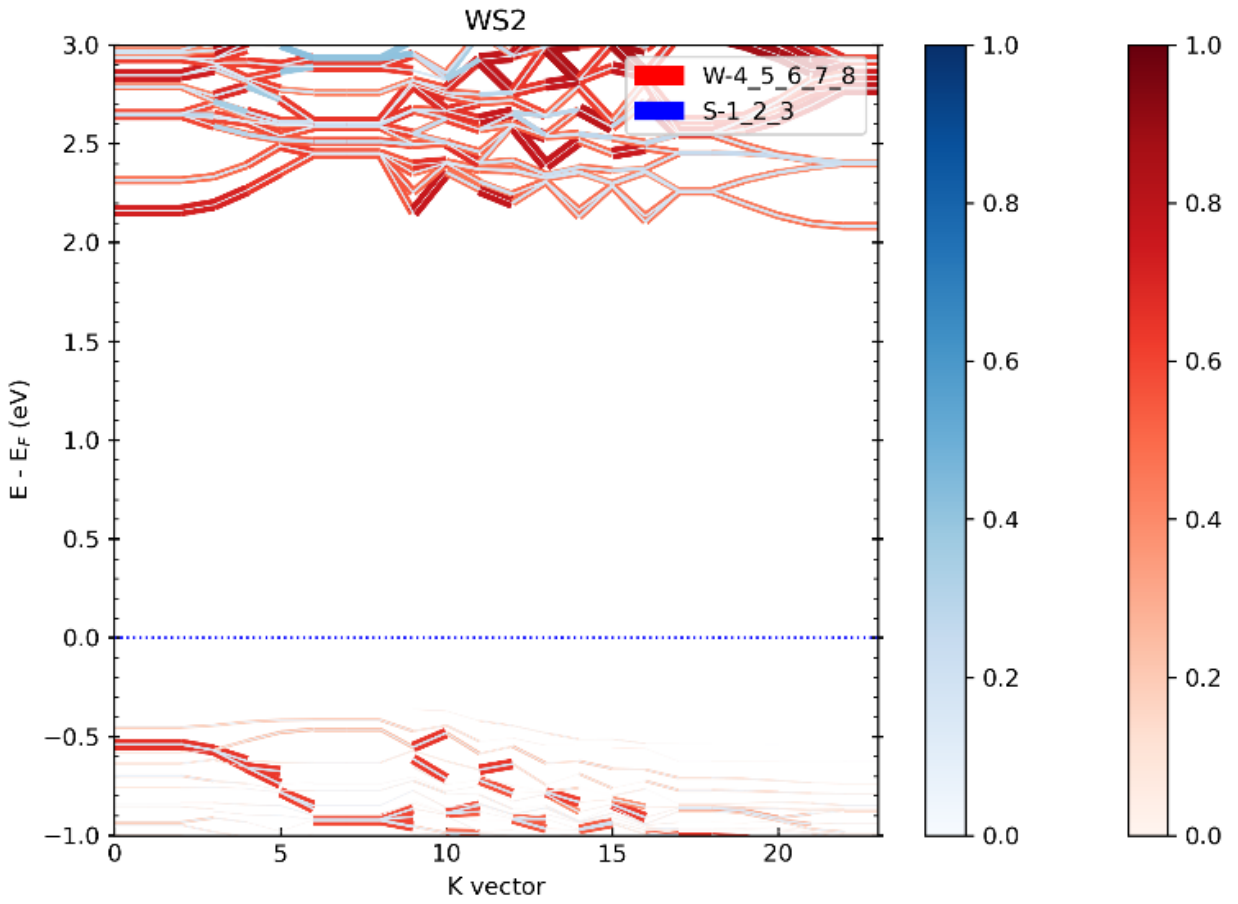


Figure S10. Projected band structure of $WS_2$/ZnO with $V_{Zn}$ bulk

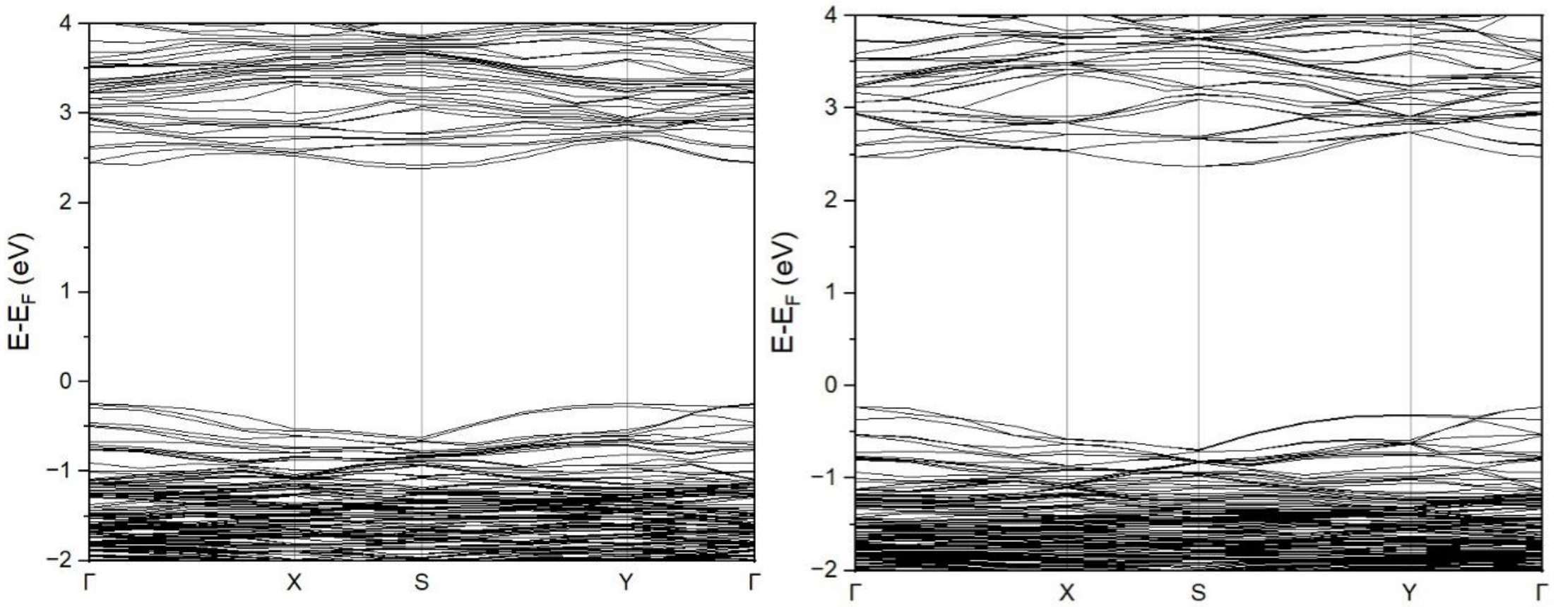


Figure S11. Band structure of $WS_2$/ZnO with $O_s$ (left) and $V_{Zn}$ 2H (right). The elimination of in-gap states due to saturation of dangling bonds is observed in comparison to $V_o$ and $V_{Zn}$.

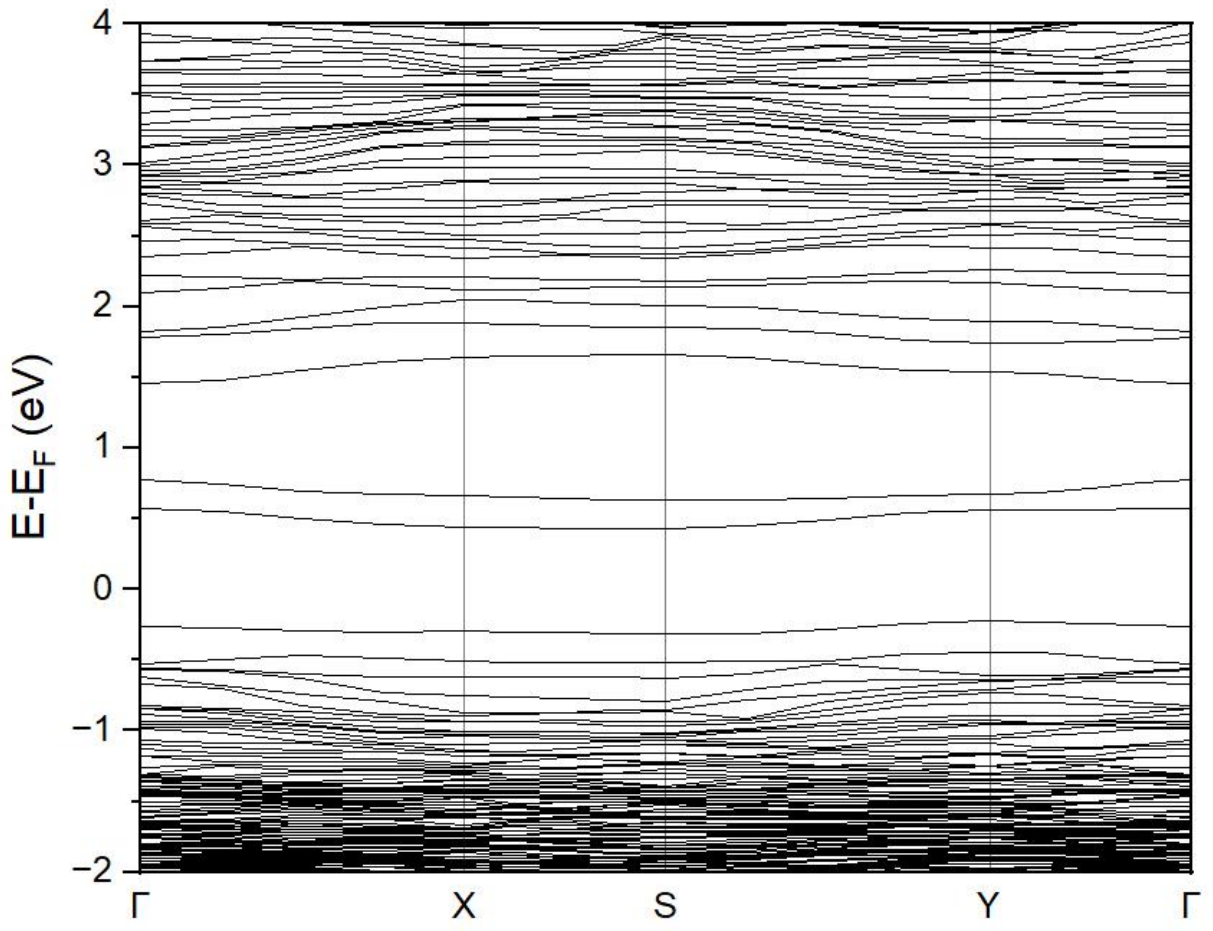


Figure S12. Band structure of the $WS_2$/ZnO heterostructure with a $V_{WS3}$ defect.

**Defect Pair**

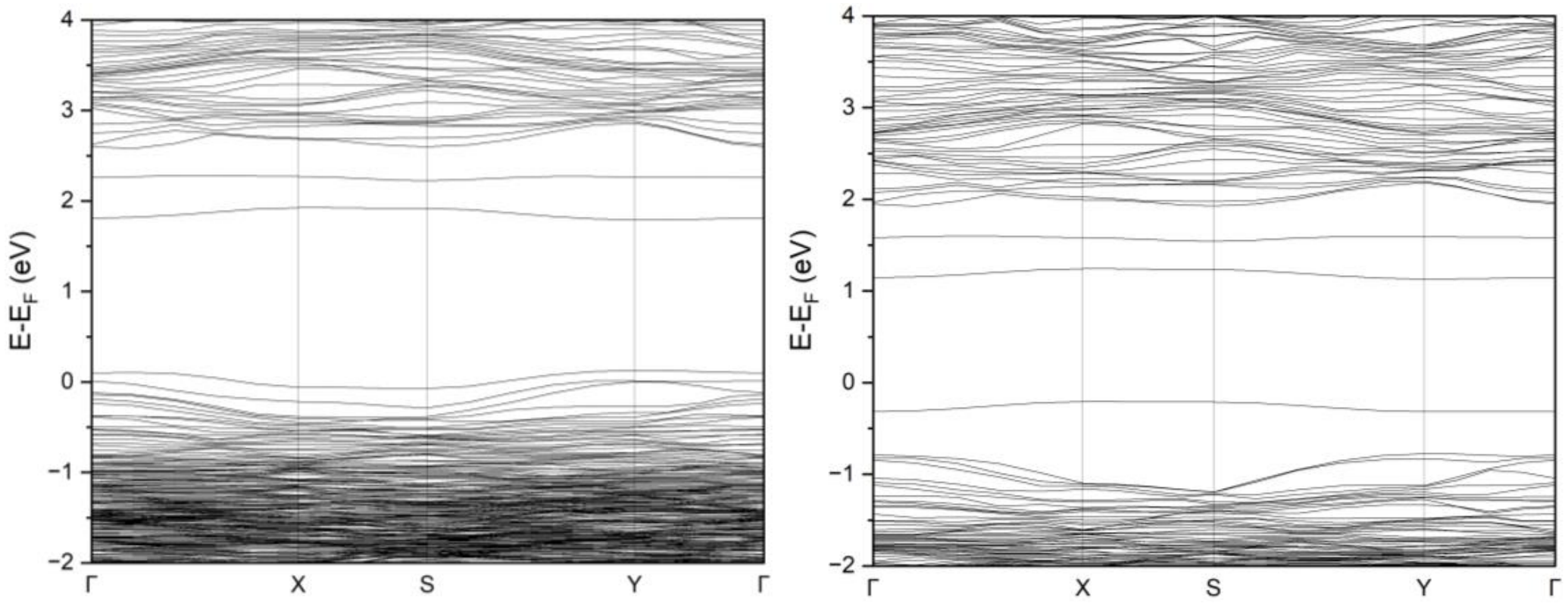


Figure S13. Band structure of $WS_2$/ZnO with defect pair: $V_s$ - $V_{Zn}$ inter and $V_s$ - $V_o$ inter

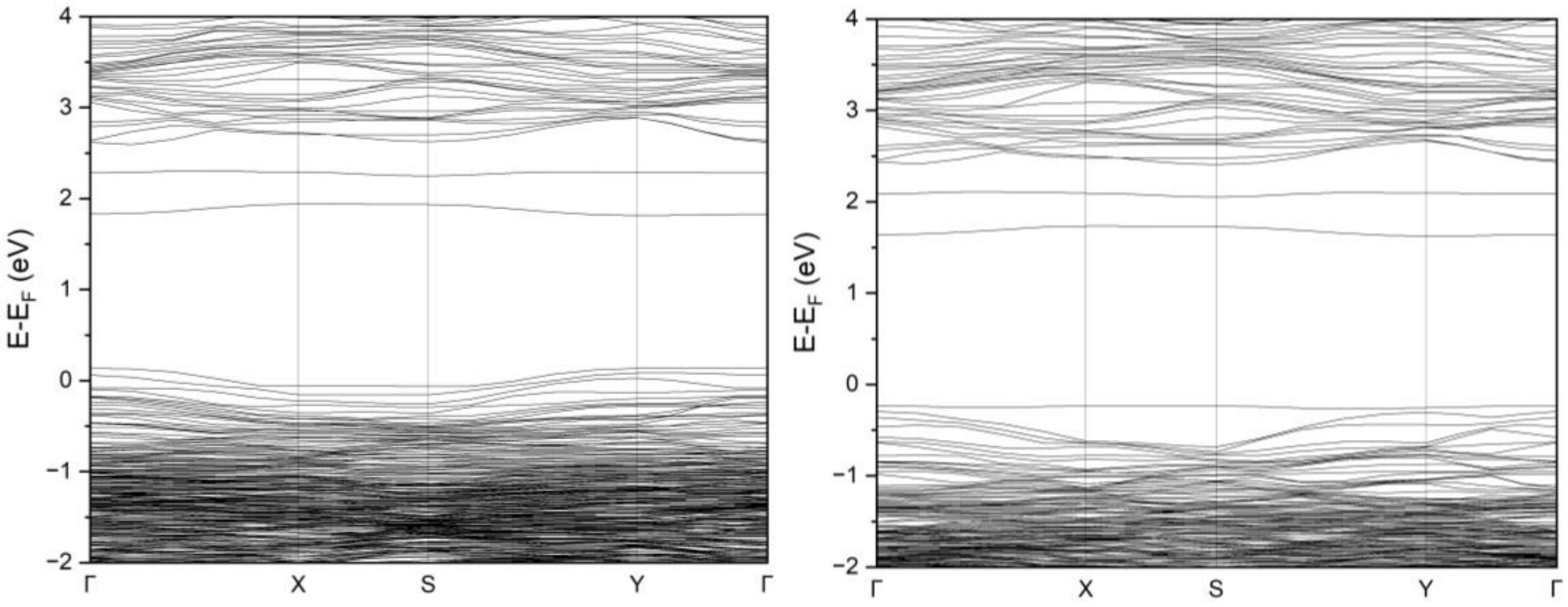


Figure S14. Band structure of $WS_2$/ZnO with defect pair: $V_s$ - $V_{Zn}$ bulk and $V_s$ - $V_o$ bulk

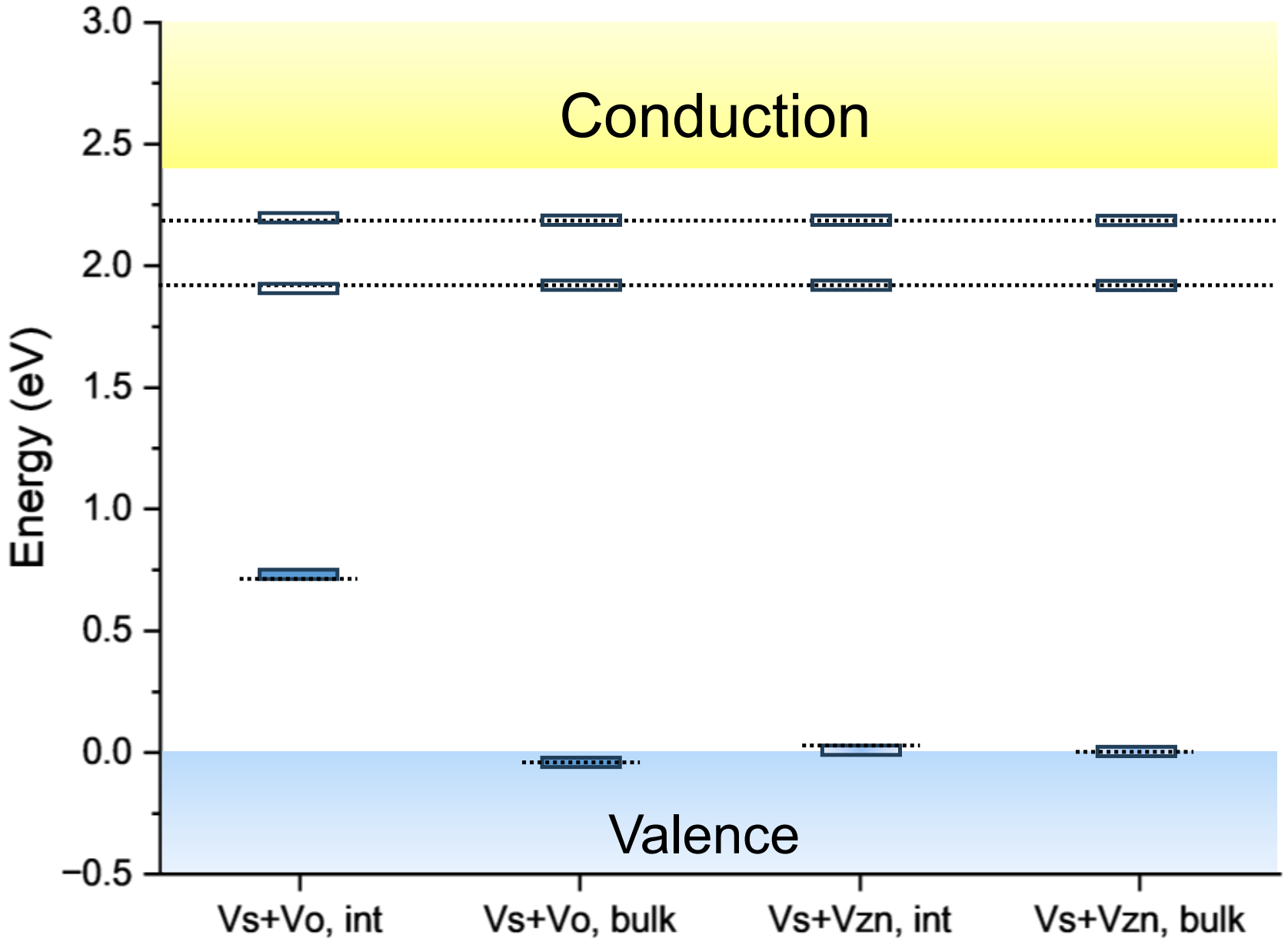


Figure S15. Electronic levels (generalized Kohn-Sham scheme) of the defect pairs, where dashed lines indicate the level position of corresponding isolated point defects for comparison

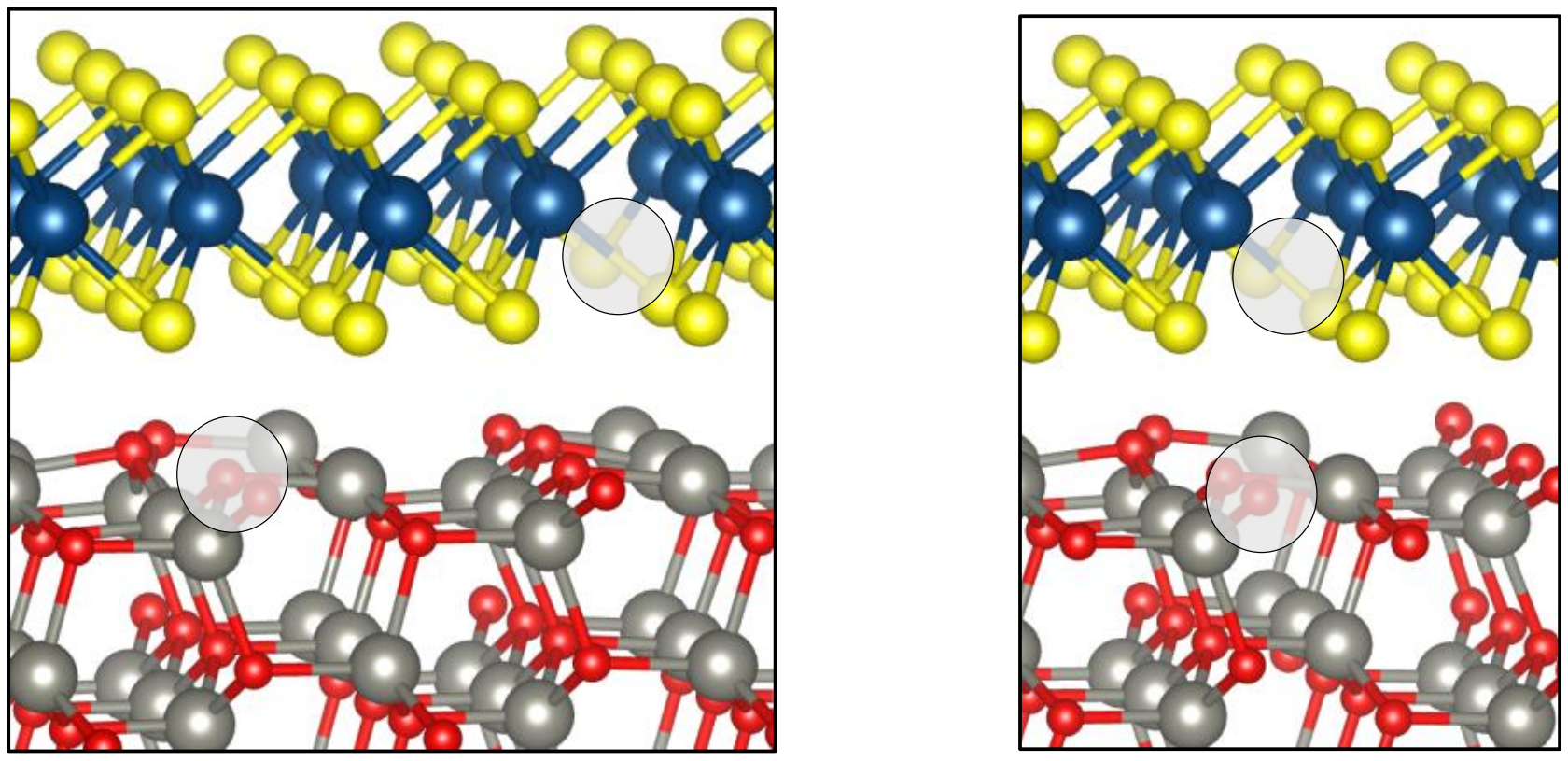

Figure S16. Comparison of two structures of $WS_2$/ZnO with $V_s$ - $V_{Zn}$ inter pair with adjacent (right) and separated (left) position.

## Binding Energy Calculations

The binding energy ($E_b$) for the dual-vacancy pair ($V_{\text{S}} + V_X$), where $V_X \in \{V_{\text{O}}, V_{\text{Zn}}\}$, is defined as the difference between the formation energy of the combined defect pair in the supercell and the sum of the formation energies of the isolated point defects:

$$E_b = E_f(V_{\text{S}} + V_X) - [E_f(V_{\text{S}}) + E_f(V_X)] \tag{1}$$

The formation energy of a defect $X$ in charge state $q$ is given by:

$$E_f(X^q) = E_{\text{tot}}(X^q) - E_{\text{tot}}(\text{pristine}) - \sum_i n_i \mu_i + q(E_V + E_F) + E_{\text{corr}}(X^q) \tag{2}$$

where $E_{\text{tot}}(X^q)$ and $E_{\text{tot}}(\text{pristine})$ are the total energies of the supercell with and without the defect, $n_i$ and $\mu_i$ are the number and chemical potential of added or removed species $i$, $E_F$ is the Fermi level referenced to the valence band maximum $E_V$, and $E_{\text{corr}}$ accounts for finite-size electrostatic charge corrections.

### a. Neutral Defect Pair ($q = 0$)

For the neutral dual vacancy ($V_{\text{S}}^0 + V_X^0$), the charge-dependent terms $q(E_V + E_F)$ and $E_{\text{corr}}$ vanish. Expanding the formation energies yields:

$$\begin{aligned} E_f(V_{\text{S}}^0 + V_X^0) &= E_{\text{tot}}(V_{\text{S}}^0 + V_X^0) - E_{\text{tot}}(\text{pristine}) + \mu_{\text{S}} + \mu_X \\ E_f(V_{\text{S}}^0) &= E_{\text{tot}}(V_{\text{S}}^0) - E_{\text{tot}}(\text{pristine}) + \mu_{\text{S}} \\ E_f(V_X^0) &= E_{\text{tot}}(V_X^0) - E_{\text{tot}}(\text{pristine}) + \mu_X \end{aligned}$$

Substituting these expressions into the binding energy definition gives:

$$\begin{aligned} E_b^0 = &\left[E_{\text{tot}}(V_{\text{S}}^0 + V_X^0) - E_{\text{tot}}(\text{pristine}) + \mu_{\text{S}} + \mu_X\right] \\ &- \left[E_{\text{tot}}(V_{\text{S}}^0) - E_{\text{tot}}(\text{pristine}) + \mu_{\text{S}}\right] \\ &- \left[E_{\text{tot}}(V_X^0) - E_{\text{tot}}(\text{pristine}) + \mu_X\right] \end{aligned}$$

The chemical potentials ($\mu_{\text{S}}, \mu_X$) cancel, yielding:

$$E_b^0 = E_{\text{tot}}(V_{\text{S}}^0 + V_X^0) + E_{\text{tot}}(\text{pristine}) - E_{\text{tot}}(V_{\text{S}}^0) - E_{\text{tot}}(V_X^0) \tag{3}$$

### b. Charged Defect Pairs

Near the conduction band minimum (CBM), the isolated sulfur vacancy adopts a $-2$ charge state ($V_{\text{S}}^{-2}$). For the oxygen vacancy, $V_{\text{O}}$ remains neutral ($V_{\text{O}}^0$), giving a combined $-2$ defect pair ($V_{\text{S}}^{-2} + V_{\text{O}}^0$). For the zinc vacancy, $V_{\text{Zn}}$ adopts a $-2$ charge state ($V_{\text{Zn}}^{-2}$), yielding a combined $-4$ defect pair ($V_{\text{S}}^{-2} + V_{\text{Zn}}^{-2}$).

In general, for a charged defect pair ($V_{\text{S}}^{q_1} + V_X^{q_2}$) with total charge $q_{\text{tot}} = q_1 + q_2$, expanding the formation energies gives:

$$\begin{aligned} E_f(V_{\text{S}}^{q_1} + V_X^{q_2}) &= E_{\text{tot}}(V_{\text{S}}^{q_1} + V_X^{q_2}) - E_{\text{tot}}(\text{pristine}) + \mu_{\text{S}} + \mu_X + q_{\text{tot}}(E_V + E_F) + E_{\text{corr}}(V_{\text{S}}^{q_1} + V_X^{q_2}) \\ E_f(V_{\text{S}}^{q_1}) &= E_{\text{tot}}(V_{\text{S}}^{q_1}) - E_{\text{tot}}(\text{pristine}) + \mu_{\text{S}} + q_1(E_V + E_F) + E_{\text{corr}}(V_{\text{S}}^{q_1}) \\ E_f(V_X^{q_2}) &= E_{\text{tot}}(V_X^{q_2}) - E_{\text{tot}}(\text{pristine}) + \mu_X + q_2(E_V + E_F) + E_{\text{corr}}(V_X^{q_2}) \end{aligned}$$

Substituting these into the binding energy expression leads to complete cancellation of the chemical potentials $\mu_i$ and Fermi level terms, since $q_{\text{tot}} - (q_1 + q_2) = 0$. The binding energy simplifies to:

$$E_b^{q_{\text{tot}}} = E_{\text{tot}}(V_{\text{S}}^{q_1} + V_X^{q_2}) + E_{\text{tot}}(\text{pristine}) - E_{\text{tot}}(V_{\text{S}}^{q_1}) - E_{\text{tot}}(V_X^{q_2}) + \Delta E_{\text{corr}} \tag{4}$$

where the net finite-size charge correction is given by:

$$\Delta E_{\text{corr}} = E_{\text{corr}}(V_{\text{S}}^{q_1} + V_X^{q_2}) - [E_{\text{corr}}(V_{\text{S}}^{q_1}) + E_{\text{corr}}(V_X^{q_2})] \tag{5}$$


\end{document}